\documentclass[twocolumn,preprintnumbers,amsmath,amssymb]{revtex4-1}

\usepackage{graphicx}% Include figure files
\usepackage{amsmath}
\usepackage{algorithm}
\usepackage{algorithmic}

\newcommand{\commentoutA}[1]{}

\begin{document}

\preprint{LA-UR-22-29595}

\title{Shadow Energy Functionals and Potentials in Born-Oppenheimer Molecular Dynamics}
%\title{Backward Error Analaysis Approach to Born-Oppenheimer Molecular Dynamics}

\author{Anders M. N. Niklasson}
\email{amn@lanl.gov}
\author{Christian F. A. Negre}
\affiliation{Theoretical Division, Los Alamos National Laboratory, Los Alamos, New Mexico 87545}

\date{\today}

\begin{abstract}
In Born-Oppenheimer molecular dynamics (BOMD) simulations based on density functional theory (DFT), the potential energy and the interatomic forces are calculated from an electronic ground state density that is determined by an iterative self-consistent field optimization  procedure, which in practice never is fully converged.
The calculated energies and the forces are therefore only approximate, which may lead to an unphysical energy drift and instabilities. Here we discuss an alternative {\em shadow} BOMD approach that is based on a backward error analysis.
Instead of calculating {\em approximate} solutions for an underlying {\em exact regular} Born-Oppenheimer potential, we do the opposite. Instead, we calculate the {\em exact} electron density, energies and forces, but for an underlying {\em approximate shadow} BO potential energy surface. In this way the calculated forces are
conservative with respect to the approximate shadow potential and generate accurate molecular trajectories with long-term energy stability. 
We show how such shadow BO potentials can be constructed at different levels of accuracy as a function of the integration time step, $\delta t$, from the constrained minimization of a sequence of systematically improvable, but approximate, shadow energy density functionals. For each energy functional there is a corresponding ground state BO potential.
These pairs of shadow energy functionals and potentials are higher-level generalizations of the original ``$0$th-level'' shadow energy functionals and potentials used in extended Lagrangian BOMD [Eur.\ Phys.\ J.\ B {\bf 94}, 164 (2021)]. The proposed shadow energy functionals and potentials are useful only within this extended dynamical framework, where also the electronic degrees of freedom are propagated as dynamical field variables together with the atomic positions and velocities.   
The theory is quite general and can be applied to MD simulations using approximate DFT, Hartree-Fock or semi-empirical methods, as well as to coarse-grained flexible charge models.
\end{abstract}

\maketitle

\section{Introduction}

The general notion of a {\em shadow} molecular dynamics provides a highly powerful concept that helps us understand and design accurate and computationally efficient simulation schemes
\cite{HYoshida90,CGrebogi90,SToxvaerd94,GJason00,ShadowHamiltonian,SToxvaerd12,KDHammonds20}.
The idea behind shadow molecular dynamics is based on a backward error analysis.
Instead of calculating approximate forces and energies for an underlying exact potential energy surface, it is often easier to calculate exact forces and energies, but for an underlying approximate {\em shadow} potential (or shadow Hamiltonian). In this way important physical properties of the simulated shadow dynamics such as time-reversibility, the conservation of the total energy and the phase-space area, can be fulfilled, because the forces of the shadow dynamics can be generated exactly.
In practice, shadow dynamics simulation methods are therefore often both more accurate and computationally more efficient compared to alternative techniques. In particular, their long-term accuracy and stability are often superior.

The shadow dynamics terminology was originally introduced in the analysis and explanation of the accuracy and long-term stability of symplectic or geometric integration schemes such as the velocity Verlet algorithm in terms of a shadow Hamiltonian \cite{HYoshida90,CGrebogi90,SToxvaerd94}. Here we use the notion of a shadow molecular dynamics in the slightly more general form that is associated with a backward error analysis. A shadow dynamics is then generated, for example, when rapid changes or discontinuities from cutoffs in the {\em exact} interatomic potential are smoothed out with an approximate {\em shadow} potential for which we can calculate
the exact forces and use longer integration time steps \cite{KDHammonds21,MHMuser22}. %\footnote{The related theory of pseudopotentials in electronic structure theory is another powerful application of a ``shadow'' potential that dates back to the 1930's \cite{PS11,DVanderbilt90,RMMartin04}, but pseudopotentials are here considered separate from the current discussion that focus on molecular dynamics.}

%\subsection{Shadow extended Lagrangian dynamics}

Shadow molecular dynamics was originally introduced in the context of classical molecular mechanics. More recently, the concept of a shadow dynamics has been applied also to non-linear self-consistent field (SCF) theory in quantum-mechanical Born-Oppenheimer molecular dynamics (QMD) simulations based on extended Lagrangian Born-Oppenheimer molecular dynamics (XL-BOMD)
\cite{ANiklasson07,ANiklasson08,MCawkwell12,JHutter12,LLin14,PSouvatzis14,ANiklasson17,ANiklasson21b}. %The electronic ground state is given from an iterative SCF solution of a set of non-linear quantum-mechanical single-particle eigenvalue equations \cite{DMarx00,ANiklasson21b} that are derived, for example, from Hartree-Fock or density functional theory (DFT) \cite{Roothaan,McWeenyHF,NMermin63,hohen,KohnSham65,NMermin65,RParr89,RMDreizler90,EEngel11}. 
The idea of a shadow molecular dynamics has been applied also to the non-linear time-dependent dynamics of superfluidity \cite{PHenning21}, as well as to flexible charge equilibration models \cite{ANiklasson21,ANiklasson21b}.
%The ability to use a shadow dynamics approach to simulations based on non-linear self-consistent field theory does not only improve the accuracy and long-term energy stability, but it can also reduce the computational cost by removing the overhead associated with the iterative self-consistent field optimization.

In this article we will revisit the construction of the approximate shadow energy functionals and potentials used in XL-BOMD simulations and show how their accuracy can be systematically improved to higher-orders  {\em as a function of the integration time step, $\delta t$}.
It is important to note that these shadow energy functionals and potentials are designed and useful only as parts of molecular dynamics simulations within the framework of XL-BOMD, where also the electronic degrees of freedom are propagated as extended dynamical variables together with the atomic positions and velocities. The interatomic forces calculated from the gradients of the shadow Born-Oppenheimer potential are exact only in this dynamical setting. For static, non-dynamical systems, the corresponding interatomic forces are only approximate, and in general not even very accurate.

%For example, in a static calculation based on Kohn-Sham density functional theory (DFT) \cite{KohnSham65}, the Kohn-Sham ground-state shadow Born-Oppenheimer potential designed for XL-BOMD corresponds to an alternative form of the Harris-Foulkes energy density functional \cite{JHarris85,MFoulkes89}. The Harris-Foulkes functional can give very good estimates of the electronic Kohn-Sham energy for approximate electron densities, but it is not suitable for gradient calculations required to evaluate the interatomic forces in a molecular dynamics simulation. The shadow energy functionals and Born-Oppenheimer potentials that we will present in this article are therefore only useful in dynamical simulations combined with XL-BOMD.

%, where the electronic degrees of freedom appear as extended dynamical field variables. Static calculations would still require a regular iterative SCF optimization procedure.

In regular QMD simulations \cite{DMarx00,MTuckerman10,ANiklasson21b} the Born-Oppenheimer potential and the interatomic forces are calculated on-the-fly from the ground-state electronic structure, which is determined from an iterative SCF optimization of some constrained non-linear energy functional, that is given, for example, from Hartree-Fock or DFT
\cite{Roothaan,McWeenyHF,NMermin63,hohen,KohnSham65,NMermin65,RParr89,RMDreizler90,EEngel11}.
In practice the iterative SCF optimization is never fully converged and always approximate.
This may create small errors in the ground state electron density, but these small errors can break time-reversibility and lead to non-conservative forces. Accumulated over time, the small errors from the approximate SCF optimization will therefore become significant. Often the errors appear as an unphysical systematic drift in the total energy, where the incompletely converged electronic structure behaves as an artificial heat source or sink \cite{DRemler90,PPulay04,JMHerbert05,ANiklasson06,TDKuhne07}, which invalidates the QMD simulations.
In the more recent formulations of XL-BOMD the shadow Born-Oppenheimer potential is designed to avoid the computational overhead and convergence errors in the iterative SCF optimization. 

In XL-BOMD the iterative SCF optimization procedure is avoided by including the electronic degrees of freedom as extended dynamical variables, in the spirit of Car-Parrinello molecular dynamics \cite{RCar85,ANiklasson21b}, in addition to the atomic positions and velocities. However, in contrast to Car-Parrinello molecular dynamics, a constrained optimization is still required to calculate the exact electronic ground state, but the optimization is performed for an approximate shadow energy functional. This optimization can be performed exactly in a single step and no iterative process is needed.
The ground state energy then defines the shadow Born-Oppenheimer potential and the corresponding conservative forces. The ability of XL-BOMD to avoid an iterative optimization and still generate exact conservative forces is thus of great practical interest, both by reducing the computational cost and by improving the accuracy and long-term stability of the molecular dynamics simulations. 

%This is achieved by including the electronic degrees of freedom as extended dynamical variables, apart from the atomic position and velocities, where the electronic degrees of freedom evolve adiabatically through a harmonic oscillator that is centered around the exact SCF ground state.

%In Hartree-Fock or density functional theory,
%the relaxed electronic ground state is given from an iterative self-consistent field
%solution of a set of non-linear eigenvalue equations, which in practice never is fully
%converged and always approximate. These errors lead to non-conservative forces with broken
%time-reversibility that often cause a significant
%systematic drift in the total energy \cite{DRemler90,PPulay04,ANiklasson07}.

%The shadow Born-Oppenheimer potential in XL-BOMD based on DFT is constructed from a constrained minimization with respect to the electron density of an approximate {\em linearized} shadow energy functional. The linearization avoids the iterative self-consistent field procedure required to find the electronic ground state of the original non-linear energy functional. Instead, the exact electronic ground state of the shadow energy functional, which defines the corresponding shadow Born-Oppenheimer potential, can be calculated in a single step directly. In this way we can not only calculate the exact forces for the shadow potential, but we also avoid the computational overhead of the iterative self-consistent field solver that is required for the non-linear energy functional.

The shadow potential approximates the exact fully-converged regular Born-Oppenheimer potential.
In the original shadow potential formulation of XL-BOMD, the error in the  forces and the potential energies scale with the size of the integration time step, $\delta t$, to the second, ${\cal O}(\delta t^2)$,  and fourth order, ${\cal O}(\delta t^4)$, respectively.
However, there seems to be no way to improve the order of the scaling. The only way to boost the accuracy is to reduce the size of the integration time step.
Here we will show how higher-levels of accuracy in the forces and shadow Born-Oppenheimer potentials can be achieved from the constrained minimization of a sequence of systematically improvable, but approximate, shadow energy functionals. For each energy functional there is a corresponding ground state Born-Oppenheimer potential. The accuracies of these pairs of shadow functionals and potentials are determined by the size of the integration time step, $\delta t$. The increased level of accuracy is thus meaningful only in the context of molecular dynamics simulations.

A higher-order accuracy in the shadow Born-Oppenheimer potential will often improve the long-term stability of a QMD simulation. 
This is of particular interest in QMD simulations of chemical systems that may have unsteady charge solutions or chemical reactions, for example, where the electronic energy gap between the Highest Occupied Molecular Orbital (HOMO) and the Lowest Unoccupied Molecular Orbital (LUMO) is opening and closing along the molecular trajectories. 

The theory will be explained in terms of general Hohenberg-Kohn DFT \cite{hohen,KohnSham65,NMermin65,RParr89,RMDreizler90,EEngel11} and is applicable to a broad range of methods, including Hartree-Fock and Kohn-Sham DFT \cite{Roothaan,McWeenyHF,NMermin63,hohen,KohnSham65,NMermin65,RParr89,RMDreizler90,EEngel11}, approximate DFT and semi-empirical methods \cite{MElstner98,MFinnis98,BHourahine20,MDewar77,MDewar85,JStewart13,CBannwarth18,PDral19,WMalone20,ZGuoqing20,CBannwarth20}, as well as to various coarse-grained polarizable charge equilibration models \cite{FJVesely77,MSprik88,WJMortier86,AKRappe91, GLamoureaux03,TVerstraelen13,SNaserifar17,DMYork96,GTabacchi02,ANiklasson21}.

The article is outlined as follows. First we review the construction of the  ``0th-level'' shadow energy functional and Born-Oppenheimer potential used in the original shadow potential formulation of XL-BOMD. We then describe how an improved ``1st-level'' pair of shadow energy functional and Born-Oppenheimer potential can be constructed.
We then derive the equations of motion in an adiabatic limit, where we assume that the extended electronic motion is rapid compared to the slower moving nuclei. This is consistent with the underlying Born-Oppenheimer approximation.
Thereafter we discuss generalizations to higher $m$th-level pairs of shadow energy functionals and potentials. The integration of the equations of motion for the electronic degrees of freedom is then explained, where we use a low-rank preconditioned Krylov subspace approximation. To better understand the shadow functionals and potentials we consider the relationship to the Harris-Foulkes functional \cite{JHarris85,MFoulkes89} in the static non-dynamical case for Kohn-Sham DFT.
We also apply our theory for the $1$st-level pairs of shadow energy functionals and potentials to a simple flexible charge equilibration model that corresponds to an orbital-free coarse-grained DFT. 
Thereafter, to summarize the results,  we present a pseudocode for XL-BOMD simulations using a 1st-level shadow energy functional and potential. We demonstrate the improved scaling and ability to treat unstable chemical systems using the 1st-level shadow energy functional and Born-Oppenheimer potential based on self-consistent charge density functional tight-binding (SCC-DFTB) theory  \cite{WHarrison80,MFoulkes89,DPorezag95,MElstner98,MFinnis98,TFrauenheim00,PKoskinen09,MGaus11,BAradi15,BHourahine20}.  
At the end we give a brief summary and our conclusions.

\section{Generalized Shadow Functionals and Potentials}

To present the pairs of energy functionals and Born-Oppenheimer potentials we will use Hohenberg-Kohn density functional theory \cite{hohen}. The corresponding Kohn-Sham expressions are generated by replacing the universal energy functional with its orbital-dependent Kohn-Sham energy functional \cite{KohnSham65}.
Generalization to Hartree-Fock theory and semi-empirical methods, as well as to coarse-grained orbital-free flexible charge models, should be straightforward \cite{ANiklasson21b}.

\subsection{Born-Oppenheimer Potential}

In Hohenberg-Kohn DFT \cite{hohen,RParr89,RMDreizler90,EEngel11},
the relaxed ground state electron density, $\rho_{\rm min}({\bf r})$, is given from a constrained minimization
of an energy density functional, $E[{\bf R},\rho]$,
over all physically relevant electron densities, $\rho$, \footnote{All physically relevant electron densities determined by anti-symmetric electron wavefunctions \cite{RMDreizler90,EEngel11}}
that integrates to the total number of electrons, $N_e$, i.e.\
\begin{equation}\label{Emin}
\rho_{\rm min}({\bf r})  = \arg \min_{\rho} \left\{E[{\bf R},\rho] \left \vert \int \rho({\bf r})d{\bf r} = N_e \right.  \right\}.
\end{equation}
The DFT energy functional, 
\begin{equation}\label{EFunc}
E[\rho] \equiv E[{\bf R},\rho] = F[\rho] + \int V_{\rm ext}({\bf R,r}) \rho({\bf r}) d{\bf r},
\end{equation}
includes a system-independent, non-linear, universal electron functional, $F[\rho]$, and an energy term with an external potential, $V_{\rm ext}({\bf R,r})$, which we here assume is from ions at the atomic positions, ${\bf R} = \{{\bf R}_I\}$.
The universal energy functional, $F[\rho]$, includes all the electron-electron interactions and the kinetic energy term. To keep it general, we may also assume ensemble generalizations where $F[\rho]$ accounts for thermal effects, including the entropy contribution at finite electronic temperatures \cite{NMermin63,NMermin65,RParr89,EEngel11,SPittalis11,APJones14,ANiklasson21b}. In the corresponding Kohn-Sham DFT the thermal effects introduces fractional occupation numbers of the Kohn-Sham orbitals
\cite{NMermin65,RParr89}, which is important to be able to describe, for example, metallic systems at finite temperatures and to stabilize calculations of systems with a small or vanishing electronic energy gap. 
%In principle, accounting for thermal effects with fractional occupation numbers goes beyond the original $v$-representability condition of Hohenberg-Kohn theory \cite{EEngel11,APJones14}, but for our presentation here this is of no practical concern. 
%We then simply assume a more general representability condition \cite{EEngel11}.

In the Born-Oppenheimer approximation  \cite{WHeitler27,MBorn27,DMarx00,MTuckerman02}
the Born-Oppenheimer potential energy surface, $U({\bf R})$, 
is determined for the fully relaxed electronic ground state, i.e.\
\begin{equation}
U({\bf R}) = E[\rho_{\rm min}] + V_{\rm nn}({\bf R}),
\end{equation}
which includes the additional ion-ion repulsion energy term, $V_{\rm nn}({\bf R})$.
The motion of the atoms can then be generated by integrating Newton's equation of motion,
\begin{equation}
M_I {\bf \ddot R}_I = - \nabla_I U({\bf R}),
\end{equation}
where $\{M_I\}$ are the atomic masses, one for each atom $I$, and the dots denote the time derivatives. 

In general, the calculation of the ground state density, $\rho_{\rm min}({\bf r})$, requires
some form of iterative optimization procedure or SCF approach, because of the non-linearity of the universal energy functional, $F[\rho]$. For example, in Kohn-Sham DFT the SCF optimization requires repeated diagonalizations of the effective single-particle Kohn-Sham Hamiltonians. This can cause a significant computational overhead and in practice the solution is never fully converged and only approximate. Force terms that in general are very difficult, if not impossible to calculate in practice, like 
\begin{align} 
&\int \left(\delta E[\rho]\big / \delta \rho({\bf r
})\right)\left(\partial \rho({\bf r}) \big / \partial {\bf R}_I \right)\big \vert_{\rho \approx \rho_{\rm min}} d{\bf r}, \label{drho}
\end{align}
are therefore not vanishing exactly, because $\left(\delta E[\rho]\big / \delta \rho({\bf r})\right)$ is vanishing only if $\rho({\bf r}) = \rho_{\rm min}({\bf r})$ 
\footnote{Notice that exact ground state, $\rho_{\rm min}({\bf r})$, and its approximate ground state solution, $\rho \approx \rho_{\rm min}$, depend on ${\bf R}$, i.e.\  $\rho_{\rm min}({\bf r}) \equiv \rho_{\rm min}({\bf R},{\bf r})$, but we have dropped the explicit ${\bf R}$-dependencies in our simplified notation}. Insufficiently converged solutions for the electronic ground state density, and where the non-vanishing force term in Eq.\ (\ref{drho}) is ignored, therefore lead to non-conservative forces that may invalidate a molecular dynamics simulation \cite{DRemler90,PPulay04,JMHerbert05,ANiklasson06,TDKuhne07}. Recent formulations of  XL-BOMD were developed to overcome these shortcomings \cite{ANiklasson21b}.

\subsection{Zeroth-Level Shadow Functional and Born-Oppenheimer Potential}

In the more recent formulations of XL-BOMD \cite{ANiklasson21b}, the energy functional, $E[\rho]$
in Eq.\ (\ref{EFunc}), is approximated by a linearized {\em shadow} energy functional, 
\begin{equation} \label{E0}
{\cal E}^{(0)}[\rho,n^{(0)}] = E[n^{(0)}] + \int \frac{\delta E[\rho]}{\delta \rho({\bf r})}\Big \vert_{n^{(0)}}\left(\rho({\bf r}) - n^{(0)}({\bf r}) \right)d{\bf r},
\end{equation}
which is given by a linearization of $E[\rho]$ around some approximate $0$th-level 
ground state density, $n^{(0)}({\bf r}) \approx \rho_{\rm min}({\bf r})$.
More generally, we can create a $0$th-level shadow energy functional \cite{ANiklasson21,ANiklasson21b} by some approximation, where
\begin{equation} \label{E0_general}
    {\cal E}^{(0)}[\rho,n^{(0)}] = E[n^{(0)}] + {\cal O}(\vert \rho - n^{(0)}\vert^2).
\end{equation}
This generalization is of particular interest in formulations of orbital-free flexible-charge equilibration models. It allows more freedom in the construction of the shadow energy functional, e.g.\ where parts of $E[\rho]$ are expanded to second order in $\rho$ to guarantee a unique ground state solution\cite{ANiklasson21,ANiklasson21b}.
The corresponding $n^{(0)}$-dependent ground state electron density, $\rho_{\rm min}[n^{(0)} ]$, is then given by the constrained minimization as in Eq.\ (\ref{Emin}), where
\begin{equation}\label{rho0_min}
\rho_{\rm min}[n^{(0)} ]({\bf r})  = \arg \min_{\rho} \left\{{\cal E}^{(0)} [\rho,n^{(0)}] \left \vert \int \rho({\bf r})d{\bf r} = N_e \right.  \right\}.
\end{equation}
With the minimization we here mean the lowest {\em stationary} solution over all physically relevant electron densities with $N_e$ number of electrons.
%\footnote{or more generally N-representable electron densities \cite{RMDreizler90,EEngel11}}. 
The relaxed ground state density then defines the approximate, $n^{(0)}$-dependent,
shadow Born-Oppenheimer potential,
\begin{equation}\label{U0}
{\cal U}^{(0)} ({\bf R},n^{(0)}) = {\cal E}^{(0)}\left[\rho_{\rm min}[n^{(0)}],n^{(0)}\right] + V_{\rm nn}({\bf R}).
\end{equation}
The advantage with this $0$th-level shadow energy functional, ${\cal E}^{(0)}[\rho,n^{(0)}]$  
in Eq.\ (\ref{E0}), is that the ground state density, $\rho_{\rm min}[n^{(0)} ]({\bf r})$,
can be calculated without requiring any iterative optimization procedure to find a SCF solution -- at least if we have found some appropriate shadow energy functional, ${\cal E}^{(0)}[\rho,n^{(0)}]$, consistent with Eq.\ (\ref{E0_general}). Instead, the
exact ground state electron density can be calculated directly in a single step, because all the non-linearities in $E[\rho]$ with respect to $\rho$ that would require an iterative solution have been removed in ${\cal E}^{(0)}[\rho,n^{(0)}]$.
In Kohn-Sham density functional theory, the exact minimization is reached in a single construction and diagonalization of the Kohn-Sham Hamiltonian, and in the corresponding coarse-grained charge equilibration models \cite{ANiklasson21,ANiklasson21b}
the relaxed ground state is given from the solution of a quasi-diagonal system of linear equations, which has a simple direct analytical solution.
In this way, any possible convergence problems and associated inconsistencies between the calculated
ground state density, $\rho_{\rm min}[n^{(0)} ]$, and the shadow Born-Oppenheimer
potential, ${\cal U}^{(0)} ({\bf R},n^{(0)})$, are avoided. 

Because of the linearization in the energy functional the error in the shadow Born-Oppenheimer potential is of second order in the residual function, $f[n^{(0)}]({\bf r}) = \rho_{\rm min}[n^{(0)}]({\bf r}) - n^{(0)}({\bf r})$, i.e.\
\begin{equation}
\left \vert {\cal U}^{(0)} - U \right \vert \propto \left \vert\rho_{\rm min}[n^{(0)}] - n^{(0)} \right \vert^2.
\end{equation} 
The approximate density, $n^{(0)}$, therefore needs to be close to the relaxed ground state 
density, $\rho_{\rm min}[n^{(0)}]$ or $\rho_{\rm min}$, to ensure that the error in the approximate shadow potential is small \footnote{If $n^{(0)} = \rho_{\rm min}$ then $\rho_{\rm min}[n^{(0)}] = n^{(0)}$ and ${\cal U}^{(0)} = U$.}. Below we will show how this is achieved
in QMD simulations by propagating the approximate ground state density, $n^{(0)}$, as a dynamical field variable within an extended Lagrangian formulation, where $n^{(0)} \equiv n^{(0)}({\bf r},t)$ is propagated by a harmonic oscillator
that is centered around the optimized ground state density, $\rho_{\rm min}[n^{(0)}]({\bf r})$, along the molecular trajectories. But before we present the extended Lagrangian molecular dynamics scheme we will show how the $0$th-level shadow energy functional and Born-Oppenheimer potential can be improved in accuracy.

\subsection{First-Level Shadow Functional and Born-Oppenheimer Potential}

The accuracy of the approximate $0$th-level shadow energy functional, ${\cal E}^{(0)}[\rho,n^{(0)}]$ in Eq.\ (\ref{E0}) or Eq.\ (\ref{E0_general}), can be improved.
However, a straightforward expansion of $E[\rho]$ to higher orders in $\rho$ would not help, because this would require some iterative solution to the constrained minimization problem of a non-linear energy functional.
Instead, we have to improve the accuracy of the approximate $0$th-level energy functional without loosing the linearity in $\rho$.
We can achieve this by improving the estimate of $n^{(0)}$
to be even closer to the exact ground state density, $\rho_{\rm min}$, in Eq.\ (\ref{Emin}).
This can be accomplished with an updated and more accurate density, $n^{(1)}({\bf r})$, which is given by a single Newton optimization step (for multiple steps see \ref{ManyNewton}),
\begin{equation}\label{Newton}\begin{array}{l}
{\displaystyle n^{(1)}({\bf r}) \equiv n^{(1)}[n^{(0)}]({\bf r}) = n^{(0)}({\bf r}) }\\
~~\\
{\displaystyle ~~~~~ - \int K^{(0)}({\bf r,r'})\left(\rho_{\rm min}[n^{(0)}]({\bf r'}) - n^{(0)}({\bf r'}) \right)d{\bf r'}},
\end{array}
\end{equation}
where the kernel $K^{(0)}({\bf r,r'})$ is the inverse Jacobian of the residual function,
$f[n^{(0)}]({\bf r}) = \rho_{\rm min}[n^{(0)}]({\bf r}) - n^{(0)}({\bf r})$. This means that
\begin{equation} \label{K0}
\int K^{(0)}({\bf r,r'}) \frac{\delta \left(\rho_{\rm min}[n^{(0)}]({\bf r'}) - n^{(0)}({\bf r'}) \right)}{{\delta n^{(0)}({\bf r''})}}d{\bf r'} = \delta({\bf r-r''}).
\end{equation}

The Newton step in Eq.\ (\ref{Newton}) (under reasonable conditions) is quadratically convergent such that
\begin{equation}
\vert \rho_{\rm min} -n^{(1)}\vert \propto  \vert  \rho_{\rm min} - n^{(0)}  \vert ^2 \propto \vert  \rho_{\rm min}[n^{(0)}] - n^{(0)}  \vert ^2.
\end{equation}
We here assume that the functional is sufficiently well-behaved and that $n^{(0)}$ is close enough to the exact ground state density, $\rho_{\rm min}$, to achieve the quadratic convergence.

The shadow energy functional can now be improved in accuarcy by using the updated density, $n^{(1)}$, instead of $n^{(0)}$ in the linearization of the energy functional. This updated and improved  approximate $1$st-level shadow energy functional is then given by
\begin{equation}\label{E1}\begin{array}{l}
{\displaystyle {\cal E}^{(1)}\left[\rho,n^{(1)}\right] = E\left[n^{(1)}\right] }\\
~~\\
{\displaystyle ~~~~ + \int \frac{\delta E\left[\rho]\right]}{\delta \rho ({\bf r})}\Big \vert_{n^{(1)}} \left(\rho({\bf r}) - n^{(1)}({\bf r}) \right)d{\bf r}},\\
\end{array}
\end{equation}
or more generally as an approximation where
\begin{align}\label{E1_general}
 E[\rho] & = {\cal E}^{(1)}\left[\rho,n^{(1)}\right] + {\cal O}( \vert \rho - n^{(1)} \vert^2).
\end{align}
The updated $n^{(1)}$-dependent ground state density is then given from the constrained minimization,
\begin{equation}\label{rho1_min}
\rho_{\rm min}[n^{(1)}]({\bf r})  = \arg \min_{\rho} \left\{{\cal E}^{(1)} \left[\rho,n^{(1)}\right] \left \vert \int \rho({\bf r}) = N_e \right.  \right\} ,
\end{equation}
with respect to variationally stationary solutions.
This ground state density defines our $1$st-level shadow Born-Oppenheimer potential,
\begin{equation}\label{U1}\begin{array}{l}
{\displaystyle {\cal U}^{(1)} ({\bf R},n^{(0)}) \equiv {\cal U}^{(1)} ({\bf R},n^{(1)}[n^{(0)}]) }\\
~~\\
{\displaystyle  = {\cal E}^{(1)}\left[\rho_{\rm min}[n^{(1)}],n^{(1)}\right] + V_{\rm nn}({\bf R}) }.
\end{array}
\end{equation}
It is important to have the $1$st-level shadow potential, ${\cal U}^{(1)} ({\bf R},n^{(0)})$, expressed as a function of $n^{(0)}$ and not of $n^{(1)}$. We can do so because $n^{(1)}$ is determined from $n^{(0)}$ in Eq.\ (\ref{Newton}), where $n^{(1)} \equiv n^{(1)}[n^{(0)}]$. We will take advantage of this relation in the next section, where $n^{(0)}$ is propagated as a dynamical field variable, $n^{(0)}({\bf r},t)$.

The constrained minimization in Eq.\ (\ref{rho1_min}) can be achieved, in general,
in a single step without requiring any iterative optimization procedure, thanks to the linear dependency of $\rho$ in the shadow energy functional, ${\cal E}^{(1)} \left[\rho,n^{(1)}\right]$. No iterative self-consistent optimization procedure is needed.

The error in the $1$st-level shadow potential scales as
\begin{equation}
 \vert {\cal U}^{(1)} - U  \vert \propto  \vert \rho_{\rm min}[n^{(1)}] -  n^{(1)} \vert^2
\propto \vert \rho_{\rm min}[n^{(0)}]-n^{(0)}  \vert^4,
\end{equation}
thanks to the quadratic convergence of the Newton update of $n^{(1)}$ in Eq.\ (\ref{Newton}), where the size of the
residual function, $f[n^{(0)}]({\bf r})=  \rho_{\rm min}[n^{(0)}]({\bf r}) - n^{(0)}({\bf r})$, decays quadratically in a single Newton step.

The Newton step in Eq.\ (\ref{Newton}) is similar to an SCF iteration step. However, in Kohn-Sham DFT, the Newton update does not require
any additiontal Hamiltonian diagonalization. In QMD simulations the Newton step can be performed using a preconditioned Krylov subspace expansion
\cite{ANiklasson20,ANiklasson20b,ANiklasson21b,VGavini22}, where each Krylov subspace vector can be determined from response calculations using quantum perturbation theory. 
%The cost of calculating a very good preconditioner can be expensive, but it QMD simulations the preconditioner can be reused, often over thousands of integration time steps before an updated preconditioner is needed. In practice the overhead is therefore quite small. 
The preconditioned Krylov subspace expansion used to approximate the kernel, $K^{(0)}$, acting on the residual function  \cite{ANiklasson20,ANiklasson20b,ANiklasson21b,VGavini22,CNegre22} is described in more detail in Sec.\ \ref{Krylov}. 

%In Kohn-Sham density functional theory we can assume
%that the Kohn-Sham Hamiltonian already has been diagonalized for the calculation
%of $\rho_{\rm min}[n^{(0)}]$. The molecular eigenstates can then be reused in the response calculations necessary for the Krylov subspace expansion. In this way the computational overhead of the Newton step in Eq.\ (\ref{Newton}) is limited. 

In practice the preconditioned Krylov subspace expansion of the kernel, $K^{(0)}$, is truncated and only approximate. The density update in Eq.\ (\ref{Newton}) is then given by a quasi-Newton step, which in general has a slower, non-quadratic convergence.

\subsection{Pairs of Shadow Functionals and Potentials}

A key concept in our presentation are
{\rm pairs} of energy functionals and Born-Oppenheimer potentials. The potential is always given from a constrained minimization over the electron density of an energy functional, where the initial pair of electronic energy functional and Born-Oppenheimer potential, corresponding to regular DFT, is given by
\begin{equation}\label{Pair_BO}
    \Big\{E[\rho], U({\bf R})\Big\}. 
\end{equation}
This pair in then replaced, first by the $n^{(0)}$-dependent $0$th-level shadow energy functional and potential,
\begin{equation}\label{Pair_S0}
    \left\{{\cal E}^{(0)}[\rho,n^{(0)}], {\cal U}^{(0)}({\bf R},n^{(0)})\right\},
\end{equation}
and then by the 
$n^{(0)}$-dependent $1$st-level shadow energy functional and potential,
\begin{equation}\label{Pair_S1}
    \left\{{\cal E}^{(1)}[\rho,n^{(0)}], {\cal U}^{(1)}({\bf R},n^{(0)})\right\}.
\end{equation}
The regular functional-potential pair in Eq.\ (\ref{Pair_BO}) is in practice difficult to represent exactly, because the calculated Born-Oppenheimer potential, $U({\bf R})$, at least in practice, is never given by the exact ground state of the energy functional, $E[\rho]$. An accurate match between $E[\rho]$ and $U({\bf R})$ can only be achieved by an expensive iterative optimization procedure, because of the non-liniarity of $E[\rho]$. This is in contrast to the $0$th-level shadow functional-potential pair in Eq.\ (\ref{Pair_S0}), which easily are matched at only a modest cost, because no iterative optimization is required. What we have presented so far is how we can construct an updated 1st-level pair of shadow energy functionals and potentials in Eq.\ (\ref{Pair_S1}) that also can be matched exactly. This higher-level generalization has an improved level of accuracy. 

\section{Extended Lagrangian Born-Oppenheimer Molecular Dynamics}

In a QMD simulation the initial approximate ground state density, $n^{(0)}({\bf r})$, around
which the linearization is performed for the construction of the shadow energy functional 
will get further and further away from the corresponding exact ground state density, 
$\rho_{\rm min}({\bf r})$, as the atoms are moving away from the initial configuration. 
The accuracy of the shadow energy functional and the corresponding shadow 
Born-Oppenheimer potential will then get successively worse.
The density, $n^{(0)}({\bf r})$, therefore needs to be updated.
One way is to update the density as a function of the atomic positions, for example, where
$n^{(0)}({\bf r}) \equiv n^{(0)}({\bf R, r}) = \sum_{I} n^{\rm atom}_I({\bf r}-{\bf R}_I)$, is the superposition of separate neutral atomic electron denisities, $\{n^{\rm atom}_I({\bf r}-{\bf R}_I)\}$, centered around the atomic positions, ${\bf R} = \{{\bf R}_I\}$.
However, this would lead to difficulties calculating forces, as in Eq.\ (\ref{drho}),
including all the density-dependent energy terms -- one for each atom.
If $n^{(0)}({\bf r})$ would be the variational ground state these terms would all vanish, but this is only true for the 
optimized densities, $\rho_{\rm min}[n^{(0)}]({\bf r})$
or $\rho_{\rm min}[n^{(1)}]({\bf r})$, with respect to the shadow potentials, 
${\cal U}^{(0)}({\bf R},n^{(0)})$ or ${\cal U}^{(1)}({\bf R},n^{(0)})$.  Even in this case partial derivatives, $\partial n^{(0)}/\partial {\bf R}_I$, would need to be calculated.
A solution to these problems is offered by XL-BOMD, where $n^{(0)}$ is propagated as a dynamical field variable, $n^{(0)}({\bf r},t)$ \cite{ANiklasson21b}.

\subsection{Extended Lagrangian}

In XL-BOMD, we include the approximate ground state density,
$n^{(0)}({\bf r})$, and its time derivative as additional dynamical field variables, $n^{(0)}({\bf r},t)$ and ${\dot n}^{(0)}({\bf r},t)$, in an extend Lagrangian
formalism, beside the nuclear positions and their velocities, ${\bf R}(t)$ and ${\bf \dot R}(t)$. The dynamics of $n^{(0)}({\bf r},t)$ is generated by an extended harmonic oscillator that is centered around the optimized ground state of the shadow potential, $\rho_{\rm min}[n^{(0)}]$, along the molecular trajectories. In this way $n^{(0)}({\bf r},t)$ closely follows the ground state such that the error in the shadow potential does not increase along the trajectory.

In the Euler-Lagrange equations of motion, the partial derivatives only appear with respect to each single dynamical variable, with all the  other dynamical variables being constant. The calculations of $n^{(0)}$-dependent force terms, e.g.\ 
\begin{align}
    &\int \frac{\delta {\cal U}^{(m)}}{\delta n^{(0)}({\bf r})} 
\frac{\partial n^{(0)}({\bf r})}{ \partial {\bf R}_I} d{\bf r}, ~~(m = 0 ~{\rm or}~1),
\end{align}
can therefore be avoided.
 Additional force terms, such as
\begin{align}
&\int \left( \frac{ \delta {\cal E}^{(m)}[\rho,n^{(m)}]}{ \delta \rho({\bf r})}\right) \left( \frac{ \partial \rho({\bf r}) }{ \partial {\bf R}_I}\right )\Big \vert_{\rho = \rho_{\rm min}[n^{(m)}]} d{\bf r},
\end{align} 
can also be ignored, 
because $\rho_{\rm min}[n^{(m)}]$ is determined from the condition that 
\begin{align}
\frac{ \delta {\cal E}^{(m)}[\rho,n^{(m)}]}{\delta \rho}\Big \vert_{\rho = \rho_{\rm min}[n^{(m)}]} = 0, ~~(m = 0~{\rm or}~1).
\end{align}
This reduces not only the computational cost, but also makes it possible to calculate ``exact'' conservative forces that generate stable long-term molecular trajectories.

We can now define the $1$st-level extended Lagrangian, ${\cal L}^{(1)}$, in XL-BOMD, using the $1$st-level shadow Born-Oppenheimer potential, where
\begin{equation}\label{XL1}\begin{array}{l}
{\displaystyle {\cal L}^{(1)}({\bf R,\dot R},n^{(0)},{\dot n^{(0)}}) = \frac{1}{2} \sum_I M_I \vert { \bf \dot R}_I\vert^2 - {\cal U}^{(1)}({\bf R},n^{(0)} )}\\
~~\\
{\displaystyle  + \frac{1}{2} \mu \int \vert {\dot n}^{(0)}({\bf r})\vert^2 d{\bf r} 
 - \frac{1}{2} \mu \omega^2 \iint \left(\rho_{\rm min}[n^{(0)}]({\bf r}) - n^{(0)}({\bf r})\right) }\\
~~\\
{\displaystyle \times T^{(0)}({\bf r,r'})\left(\rho_{\rm min}[n^{(0)}]({\bf r'}) - n^{(0)}({\bf r'})\right)d{\bf r} d{\bf r'}}.
\end{array}
\end{equation}
Here $n^{(0)}({\bf r},t)$ is treated as a dynamical field variable with its time derivative, ${\dot n}^{(0)}({\bf r},t)$,
and some chosen mass parameter, $\mu$. This is in addition to the regular dynamical variables of the atomic motion, ${\bf R}$ and ${\bf \dot R}$. The atomic masses are given by $\{M_I\}$. The frequency of the extended harmonic oscillator is set by $\omega$ and the harmonic well is centered around $\rho_{\rm min}[n^{(0)}]({\bf r})$. $T^{(0)}({\bf r,r'})$ is a symmetric positive definite metric tensor
given by the square of a kernel, $K^{(0)}({\bf r,r'})$, where
\begin{equation}
T^{(0)} ({\bf r,r'}) = \int \left(K^{(0)}({\bf r,r''})\right)^{\dagger} K^{(0)}({\bf r'',r'})d{\bf r''}.
\end{equation}
We define the kernel, $K^{(0)}({\bf r,r'})$, as the inverse Jacobian of the residual function, 
$f[n^{(0)}]({\bf r}) = \rho_{\rm min}[n^{(0)}]({\bf r}) - n^{(0)}({\bf r})$. This means that the
kernel, $K^{(0)}({\bf r,r'})$, is the same as in Eqs.\ (\ref{Newton}) and (\ref{K0}).
In this way the dynamical density variable, $n^{(0)}({\bf r},t)$, evolves as if it would oscillate
around the much closer approximation to the exact ground state, i.e.\ $n^{(1)}({\bf r})$ from the Newton update, 
compared to the more approximate, $\rho_{\rm min}[n^{(0)}]({\bf r})$.
This definition of the kernel simplifies the equations of motion that we will derive below at the same time as it also improves the accuracy of the shadow Born-Oppenheimer potential by evolving $n^{(0)}({\bf r},t)$ around a closer approximation to the exact ground state, $\rho_{\rm min}({\bf r})$.

The only difference to the original formulation of XL-BOMD is that 
the Lagrangian, ${\cal L}^{(1)}({\bf R,\dot R},n^{(0)},{\dot n^{(0)}})$, in Eq.\ (\ref{XL1}) uses
the $1$st-level shadow Born-Oppenheimer potential, ${\cal U}^{(1)}({\bf R},n^{(0)} )$, 
instead of the $0$-th level, ${\cal U}^{(0)}({\bf R},n^{(0)})$.

%\begin{equation}
%\int K^{(0)}({\bf r,r'}) \frac{\delta \left(\rho_{\rm min}[n^{(0)}]({\bf r'}) - n^{(0)}({\bf r'}) \right)}{{\delta n^{(0)}({\bf r''})}}d{\bf r'} = \delta({\bf r-r''})
%\end{equation}
%
\subsection{Equations of Motion}

The Euler-Lagranges equations for ${\cal L}^{(1)}({\bf R,\dot R},n^{(0)},{\dot n^{(0)}})$,
\begin{equation}
\frac{d}{dt}\left(\frac{\partial {\cal L}^{(1)}({\bf R,\dot R},n^{(0)},{\dot n^{(0)}}))}{\partial {\bf \dot R}_I} \right) = \frac{\partial {\cal L}^{(1)}({\bf R,\dot R},n^{(0)},{\dot n^{(0)}}))}{\partial {\bf R}_I}
\end{equation}
and
\begin{equation}
\frac{d}{dt}\left(\frac{\delta {\cal L}^{(1)}({\bf R,\dot R},n^{(0)},{\dot n^{(0)}}))}{\delta {\dot n}^{(0)}({\bf r})} \right) = \frac{\delta {\cal L}^{(1)}({\bf R,\dot R},n^{(0)},{\dot n^{(0)}}))}{\delta n^{(0)}({\bf r})}
\end{equation}
give us the equations of motion,
\begin{equation}\label{R_Mot}\begin{array}{l}
{\displaystyle M_I{\bf \ddot R}_I =  - \frac{\partial {\cal U}^{(1)}({\bf R},n^{(0)})}{\partial {\bf R}_I}\Big \vert_{n^{(0)},n^{(1)}}}\\
~~\\
{\displaystyle  ~~ - \int \frac{\delta {\cal U}^{(1)}({\bf R},n^{(0)})}{\delta n^{(1)}({\bf r})} \frac{\partial n^{(1)}({\bf r})}{\partial {\bf R}_I}\Big \vert_{n^{(0)}} d{\bf r}}\\
~~\\
{\displaystyle  ~~~ - \frac{1}{2} \mu \omega^2 \frac{\partial}{\partial {\bf R}_I} \iint \left(\rho_{\rm min}[n^{(0)}]({\bf r}) - n^{(0)}({\bf r})\right)T^{(0)}({\bf r,r'})}\\
~~\\
{\displaystyle ~~~~ \times \left(\rho_{\rm min}[n^{(0)}]({\bf r'}) - n^{(0)}({\bf r'})\right) \Big \vert_{n^{(0)}} d{\bf r} d{\bf r'}} \\
%{\displaystyle  - \int \frac{\delta {\cal U}^{(1)}}{\delta n^{(1)}({\bf r})}\frac{\partial n^{(1)}({\bf r})}{\partial {\bf R}_I} \Big \vert_{n^{(0)}}  d{\bf r}}
\end{array}\end{equation}
and
\begin{equation}\label{n_Mot}\begin{array}{l}
{\displaystyle \mu {\ddot n}^{(0)}({\bf r}) = - \frac{\delta {\cal U}^{(1)}({\bf R},n^{(0)})}{\delta n^{(0)}({\bf r})}}\\
~~\\
%{\displaystyle -\mu \omega^2 \int K^{(0)}({\bf r,r'}) \left(\rho_{\rm min}[n^{(0)}] - n^{(0)}({\bf r'}) \right)d{\bf r'}}\\
{\displaystyle - \frac{1}{2} \mu \omega^2 \frac{\delta}{\delta n^{(0)}({\bf r})} \iint \left(\rho_{\rm min}[n^{(0)}]({\bf r'}) - n^{(0)}({\bf r'})\right) }\\
~~\\
{\displaystyle ~\times T^{(0)}({\bf r',r''})\left(\rho_{\rm min}[n^{(0)}]({\bf r''}) - n^{(0)}({\bf r''})\right)d{\bf r'} d{\bf r''}}
\end{array}
\end{equation}
These equations are far from trivial to use in a QMD simulation. However, the equations of motion
are simplified if we impose an adiabatic limit, in the same way as for the original Born-Oppeheimer approximation, where we assume that the electronic
degrees of freedom are fast compared to the slower nuclear motion. To derive the equations of motion in this adiabatic limit we first assert the following frequency dependencies in the residual functions, 
\begin{equation}\label{Res0}
\left \vert \rho_{\rm min} [n^{(0)}] - n^{(0)} \right \vert  \propto \omega^{-2},
\end{equation}
and
\begin{equation}\label{Res1}
\left \vert \rho_{\rm min} [n^{(1)}] - n^{(1)} \right \vert  \propto \omega^{-4},
\end{equation}
which are assumed to be valid in the limit of $\omega \rightarrow \infty$.
These adiabatic relations are difficult to prove {\em a priori}, but they can be shown to hold {\em a posteriori} by integrating the equations of motions that have been derived under the assumptions of Eqs.\ (\ref{Res0}) and (\ref{Res1}). This will demonstrate below in Fig.\ \ref{Fig_1}. 

Using the asserted adiabatic scaling relations in Eqs.\ (\ref{Res0}) and (\ref{Res1}),
we find (under reasonable conditions) from the definition of $n^{(1)}[n^{(0)}]({\bf r})$ in Eq.\ (\ref{Newton}) that
\begin{equation}\begin{array}{l}
{\displaystyle\frac{\delta n^{(1)}[n^{(0)}]({\bf r})}{\delta n^{(0)}({\bf r''})} 
= \frac{\delta n^{(0)}({\bf r})}{\delta n^{(0)}({\bf r''})}}\\
~~\\
{\displaystyle - \int K^{(0)}({\bf r,r'}) \frac{\delta \left(\rho_{\rm min}[n^{(0)}]({\bf r'}) - n^{(0)}({\bf r'}) \right)}{{\delta n^{(0)}({\bf r''})}}d{\bf r'} }\\
~~\\
{\displaystyle - \int \frac{\delta K^{(0)}({\bf r,r'})}{\delta n({\bf r''})}  \left(\rho_{\rm min}[n^{(0)}]({\bf r'}) - n^{(0)}({\bf r'}) \right)d{\bf r'}}\\
~~\\
{\displaystyle = \delta({\bf r-r''})- \delta({\bf r-r''})}\\
~~\\
{\displaystyle - \int \frac{\delta K^{(0)}({\bf r,r'})}{\delta n({\bf r''})}  \left(\rho_{\rm min}[n^{(0)}]({\bf r'}) - n^{(0)}({\bf r'}) \right)d{\bf r'}}\\
~~\\
{\displaystyle \propto \left \vert \rho_{\rm min}[n^{(0)}]({\bf r'}) - n^{(0)}({\bf r'}) \right \vert \propto \omega^{-2}}.
\end{array}\end{equation}
Using the same assertions we also find that 
\begin{equation}
\left \vert \frac{\delta {\cal U}^{(1)}} {\delta n^{(1)}}\right \vert  \propto \left \vert \rho_{\rm min}[n^{(1)}] - n^{(1)} \right \vert \propto \omega^{-4}.
\end{equation}
This gives us
\begin{align}
\left \vert \frac{\delta {\cal U}^{(1)}} {\delta n^{(0)}}\right \vert & = \left \vert \frac{\delta {\cal U}^{(1)}} {\delta n^{(1)}}\frac{\delta n^{(1)}} {\delta n^{(0)}}  \right \vert \\
& \propto \left \vert \rho_{\rm min}[n^{(1)}] - n^{(1)} \right \vert \times \left \vert \rho_{\rm min}[n^{(0)}]- n^{(0)} \right \vert \\
& \propto \omega^{-4} \times \omega^{-2}.
\end{align}
%Normally, when we calculate the Born-Oppenheimer potential using regular a regular SCF optimization, we neither use
%the linearized energy approximations or Newton steps. This will lead to problems in a derivation of a consistent
%dynamics in an adiabatic limit where we let $\lim \omega \rightarrow \infty$.
The scaling relation in Eq.\ (\ref{Res0}) also mean that the last gradient term in Eq.\ (\ref{R_Mot})
becomes proportional to $\mu$.  The asserted scaling relations above inserted in
the equations of motion in Eqs.\ (\ref{R_Mot}) and (\ref{n_Mot}) then give us,
%\begin{equation}\begin{array}{l}
%\left \vert \nabla_I \iint \left(\rho_{\rm min}[n^{(0)}]({\bf r}) - n^{(0)}({\bf r})\right)T^{(0)}({\bf r,r'})\left(\rho_{\rm min}[n^{(0)}]({\bf r'}) - n^{(0)}({\bf r'})\right)d{\bf r} d{\bf r'}\right \vert \propto \omega^{-2}
%\end{equation}
\begin{equation}\begin{array}{l}
{\displaystyle M_I{\bf \ddot R}_I = - \frac{\partial {\cal U}^{(1)}({\bf R},n^{(0)})}{\partial {\bf R}_I}\Big \vert_{n^{(0)}} 
+ {\cal O}\left( \omega^{-4} \right) + {\cal O}\left( \mu \right)}, \\
\end{array}
\end{equation}
and
\begin{equation}\begin{array}{l}
{\displaystyle {\ddot n}^{(0)}({\bf r}) =  {\cal O}(\mu^{-1}\omega^{-6}) + {\cal O}(\omega^{-2})}\\
~~\\
{\displaystyle ~~~~ -\omega^2 \int K^{(0)}({\bf r,r'}) \left(\rho_{\rm min}[n^{(0)}]({\bf r'}) - n^{(0)}({\bf r'}) \right)d{\bf r'}},
\end{array}
\end{equation}
where we have assumed that $\delta T^{(0)}/\delta n^{(0)}$ is bounded and $\omega$-independent as $\omega \rightarrow \infty$.
We can then derive the equations of motion in the adiabatic limit, where $\omega \rightarrow \infty$
combined with the mass-zero limit $\mu \rightarrow 0$, which here is chosen such that $\mu \omega^4 \rightarrow {\rm constant}$. This is a classical analogue to the Born-Oppenheimer approximation, where we simply stick with the original Born-Oppenheimer assumption that the electronic degrees of freedom is evolving on a much faster time scale compared to a slower nuclear motion.
In this adiabatic limit we get the final equations of motion for XL-BOMD with the $1$st-level updated shadow Born-Oppenheimer potential,
\begin{align}
M_I{\bf \ddot R}_I = & -\frac{\partial}{\partial {\bf R}_I} {\cal U}^{(1)}({\bf R},n^{(0)})\Big \vert_{n^{(0)}} \label{EOM_R},\\
{\ddot n}^{(0)}({\bf r}) = & -\omega^2 \int K^{(0)}({\bf r,r'}) \left(\rho_{\rm min}[n^{(0)}]({\bf r'}) - n^{(0)}({\bf r'}) \right)d{\bf r'} \label{EOM_n} .
\end{align}
Because $n^{(0)} \equiv n^{(0)}({\bf r},t)$ is a dynamical field variable in XL-BOMD, the partial derivatives in Eq.\ (\ref{EOM_R}) with respect to the nuclear coordinates are evaluated under a constant electron density, $n^{(0)}$. Thus, even if $n^{(0)}$ is not the variationally optimized ground state density, we can still calculate the exact forces in the adiabatic equations of motion for XL-BOMD. However, this does not work for static calculations. It only works in the context of XL-BOMD, where the electronic degrees of freedom are propagated dynamically. The 1st-level updated shadow potential is thus mainly useful only in this dynamical setting.

The equations of motion, Eqs.\ (\ref{EOM_R}) and (\ref{EOM_n}), are almost identical to the original equations of motion for XL-BOMD using the
$0$th-level Born-Oppenheimer potential \cite{ANiklasson17,ANiklasson20,ANiklasson21b}.
The only difference is that we now have the $1$st-level Born-Oppenheimer potential, ${\cal U}^{(1)}({\bf R},n^{(0)})$,
instead of the original $0$th-level ${\cal U}^{(0)}({\bf R},n^{(0)})$. The error terms neglected in the adiabatic limit, where $\mu \propto \omega^{-4}$, indicates
that the error in the interatomic force term should scale as $\omega^{-4}$. This scaling will also
be demonstrated below in Fig.\ \ref{Fig_2}.

It is important to note that even if we only would use some approximation of the kernel, $K({\bf r,r'})$, the
same equations of motion, in Eqs.\ (\ref{EOM_R}) and (\ref{EOM_n}), can be derived in an adiabatic limit.
The only difference is that the adiabatic limit has to be modified such that $\mu \omega^m \rightarrow {\rm constant}$
for some value $ m \in [1,4]$, and with a modified assertion, where 
$\vert \rho_{\rm min} [n^{(1)}] - n^{(1)} \vert  \propto \omega^{-m}$ for some value of $m \in [2,4]$. 
The scaling of the errors in the forces and the potential energy will then be different and less favorable.
Of critical importance is only that we calculate the forces from the shadow potential, ${\cal U}^{(1)}$, 
defined by the optimized ground state of a shadow energy functional that has been linearized around 
some updated $n^{(0)}$-dependent density, $n^{(1)}({\bf r}) \equiv n^{(1)}[n^{(0)}]({\bf r})$, and where 
$\big \vert \delta {\cal U}^{(1)}\big/\delta n^{(1)}\big \vert  \propto \vert \rho_{\rm min} [n^{(1)}] - n^{(1)} \vert$.
Replacing the Newton update of the electron density in Eq.\ (\ref{Newton})
with an approximate quasi-Newton scheme or any other SCF-like iteration update, should therefore also work under the same conditions.
This observation may also help explain why some earlier versions of XL-BOMD \cite{ANiklasson06,ANiklasson08,PSteneteg10,MArita14,BAradi15,LDMPeters17} often works quite well, but where a few SCF cycles often were required in each time step prior to the force evaluations, while the extended electronic degrees of freedom was propagated dynamically. Our analysis here shows us why and when we can expect these initial versions of XL-BOMD to work or fail. This insight appears analogous to how, for example, solving a system of non-linear equations with some simple {\em ad hoc} mixed iterations (which often works), can be replaced by a more transparent and efficient conjugate gradient or Newton-based method.  Once we understand the theoretically more rigorous alternative we also understand why and when the {\em ad hoc} method works and how it can be improved.

The equations of motion in Eqs.\ (\ref{EOM_R}) and (\ref{EOM_n}), 
in combination with the definition
of the $1$st-level shadow energy functional, Eq.\ (\ref{E1}), and the Born-Oppenheimer potential, Eq.\ (\ref{U1}), are some of the key results of this article.

\subsection{Higher-Level Generalizations}\label{ManyNewton}

%\begin{equation}
%n^{(2)}({\bf r} ) \equiv n^{(2)}\left[ n^{(1)}[n^{(0)}]\right]({\bf r}) = n^{(1)}({\bf r}) - \int K^{(1)}({\bf r,r'})\left(\rho_{\rm min}[n^{(1)} ]({\bf r'}) - n^{(1)}({\bf r'}) \right)d{\bf r'}
%\end{equation}
%
%\begin{equation}
%{\cal E}^{(2)}\left[\rho,n^{(2)}\right] = E\left[n^{(2)}\right] + \int \frac{\delta E\left[\rho \right]}{\delta \rho({\bf r})}\Big \vert_{n^{(2)}} \left(\rho({\bf r}) - n^{(2)}({\bf r}) \right)d{\bf r} + \int V_{\rm ext}({\bf R,r}) \rho({\bf r}) d{\bf r} 
%\end{equation}
%
%\begin{equation}
%\rho_{\rm min}\left[n^{(2)}\right]({\bf r})  = \arg \min_{\rho \in v} \left\{{\cal E}^{(2)} \left[\rho,n^{(2)}\right] \left \vert \int \rho({\bf r}) = N_e \right.  \right\} 
%\end{equation}
%
%\begin{equation}
%{\cal U}^{(2)} ({\bf R},n^{(0)}) \equiv {\cal U}^{(2)} ({\bf R},n^{(2)}\left[n^{(1)}[n^{(0)}]\right])  = {\cal E}^{(2)}\left[[\rho_{\rm min}[n^{(2)}],n^{(2)}\right] + V_{\rm nn}({\bf R}) 
%\end{equation}
%
Higher $m$th-level generalizations of the pairs of shadow energy functionals and potentials can also be designed, where the approximate higher-level density approximations to the exact ground state are updated with repeated Newton steps,
\begin{equation}\begin{array}{l}
{\displaystyle n^{(m)}({\bf r}) \equiv n^{(m)}[n^{(0)}]({\bf r})}\\
~~\\
{\displaystyle \equiv n^{(m)}\left[ n^{(m-1)}[ \ldots [n^{(0)}] ]\right]({\bf r})
= n^{(m-1)}({\bf r})}\\
~~\\
{\displaystyle - \int K^{(m-1)}({\bf r,r'})\left(\rho_{\rm min}\left[n^{(m-1)} \right]({\bf r'}) - n^{(m-1)}({\bf r'}) \right)d{\bf r'}}.
\end{array}
\end{equation}
The corresponding linearized $m$th-level shadow energy density functionals are then given by,
\begin{equation}\begin{array}{l}
{\displaystyle {\cal E}^{(m)}\left[\rho,n^{(m)}\right] = E\left[n^{(m)}\right]}\\
~~\\
{\displaystyle  + \int \frac{\delta E\left[\rho\right]}{\delta \rho({\bf r})}\Big \vert_{n^{(m)}}\left(\rho({\bf r}) - n^{(m)}({\bf r}) \right)d{\bf r}}.
\end{array}
\end{equation}
The constrained electronic ground state optimization then gives us the ground state density,
\begin{equation}\begin{array}{l}
\rho_{\rm min}\left[n^{(m)} \right]({\bf r})  \\
~~\\
~~= \arg \min_{\rho } \left\{{\cal E}^{(m)} \left[\rho,n^{(m)}\right] \left \vert \int \rho({\bf r})d{\bf r} = N_e \right.  \right\} ,
\end{array}
\end{equation}
which defines the $m$th-level shadow Born-Oppenheimer potentials,
\begin{equation}\begin{array}{l}
{\cal U}^{(m)} ({\bf R},n^{(0)}) = {\cal E}^{(m)}\left[\rho_{\rm min}\left[n^{(m)}\right],n^{(m)}\right] + V_{\rm nn}({\bf R}) .
\end{array}
\end{equation}
The adiabatic equations of motion from an $m$th-level extended Lagrangian, ${\cal L}^{(m)}$, follows in the same way as above, where
\begin{align}
M_I{\bf \ddot R}_I = & - \frac{\partial}{\partial {\bf R}_I} {\cal U}^{(m)}({\bf R},n^{(0)})\Big \vert_{n^{(0)}} \label{m_EOM_R},\\
{\ddot n}^{(0)}({\bf r}) = & -\omega^2 \int K^{(0)}({\bf r,r'}) \left(\rho_{\rm min}[n^{(0)}]({\bf r'}) - n^{(0)}({\bf r'}) \right)d{\bf r'} \label{m_EOM_n} .
\end{align}
As for the $1$st-level approximation, we have used the nested dependencies of $n^{(m)}$ on $n^{(0)}$ and let the shadow potential be a functional of $n^{(0)}$.
While the above higher-order generalization is straightforward, we have found it of little value in practice, 
because the accuracy is, in general, already very high at the $0$th-level and virtually exact at the
$1$st-level. For example, we tried to show numerically that the error in the shadow potential energy surface,
which scales as $\omega^{-4}$ for ${\cal U}^{(0)}({\bf R}, n^{(0)})$ \cite{ANiklasson17}, scales
as $\omega^{-8}$ for ${\cal U}^{(1)}({\bf R}, n^{(1)})$. However, in practical simulations this scaling was not possible to observe, because the error in the $1$st-level shadow potential
for any normal integration time steps was already at machine precision and no relevant scaling could be
demonstrated. Instead, it has to be demonstrated indirectly from the scaling of $\vert \rho_{\rm min}[n^{(1)}] - n^{(1)}\vert$ from which we get the scaling of $\vert {\cal U}^{(1))} - U \vert \propto \vert \rho_{\rm min}[n^{(1)}] - n^{(1)}\vert^2$. In the following we will therefore ignore any higher-level generalizations beyond the $1$st-level.

\subsection{Integrating the electronic equation of motion}

To integrate the equations of motion for the nuclear degrees of freedom
in Eq.\ (\ref{EOM_R}) we can use a leapfrog velocity Verlet scheme, whereas the integration of the harmonic oscillator equation of motion in Eq.\ (\ref{EOM_n}) for the
extended electronic degrees of freedom requires some care. In principle,
the same Verlet integration scheme could be used also for the electronic
propagation. However, typically we need to include some weak form of dissipation 
that keeps $n^{(0)}({\bf r})$ synchronized with the trajectories of the atomic positions
and the exact Born-Oppenheimer ground state
\cite{ANiklasson09,PSteneteg10,GZheng11,AOdell09,AOdell11,ANiklasson21b}.
This modified Verlet integration scheme has the following form,
\begin{equation}\label{Int_n}
{\displaystyle {\bf n}^{(0)}_{j+1} = 2 {\bf n}^{(0)}_j - {\bf n}^{(0)}_{j-1} + \delta t^2 {\bf \ddot n}^{(0)}_j +
\alpha \sum_{l = 0}^{l_{\rm max}} c_l {\bf n}^{(0)}_{j-l}},
\end{equation}
where we use a convenient vector notation, with ${\bf n}^{(0)}_j \equiv {\bf n}^{(0)}(t_0 + j\delta t) \in {\boldmath R}^N$, $~j = 0,1,2,\ldots$~.
The first three terms on the right-hand side of Eq.\ (\ref{Int_n}) are the regular Verlet terms,
whereas the last term is an additional weak dissipative {\em ad hoc} damping force.
An optimized set of coefficients of $\alpha$, $\{c_l\}$, and the dimensionless constant $\kappa = \delta t^2 \omega^2$,
for various orders of $l_{\rm max}$ can be found in Ref.\ \cite{ANiklasson09}.
As an alternative to such modified Verlet integration schemes, we may
connect the electronic degrees of freedom to a thermostat, i.e.\ a stochastic
Langevin-like dynamics or a chained Nose-Hoover thermostat, which also keeps
the electronic degrees of freedom synchronized with the ground state solution determined by the nuclear coordinates \cite{ILeven19,DAn20}.

In the initial time step we can set all densities $\{{\bf n}_j\}$ equal to the optimized regular Born-Oppenheimer ground state density, ${\boldsymbol \rho}_{\rm min}$.

It is important to note that we always use a constant for the product $\delta t^2 \omega^2 = \kappa$
in our simulations. This means that $\delta t \propto \omega^{-1}$, as long as we use the same Verlet integration scheme with a constant size of the integration time step, $\delta t$. This controls the way we can understand the scaling and the order of the accuracy, for example, of the forces (See Fig.\ \ref{Fig_2}), as a function of the chosen size of the integration time step, $\delta t$, or the inverse frequency, $\omega^{-1}$.

\subsection{Approximating the kernel with preconditioned Krylov subspace}\label{Krylov}

In addition to the modified Verlet integration, we also need to approximate
the kernel, $K({\bf r,r'})$, both in the integration of the electronic degrees of freedom in Eq.\ (\ref{EOM_n}) 
and for the Newton update of the density, $n^{(0)}$, to $n^{(1)}$ in Eq.\ (\ref{Newton}).
The kernel is the same and it is acting on the same residual, apart from a trivial constant factor $\omega^2$. The approximation of the kernel acting on the residual therefore only needs to be performed once every integration time step. 

In the more convenient matrix-vector notation, Eq.\ (\ref{EOM_n}) or Eq.\ (\ref{m_EOM_n}) is given by
\begin{equation}
{\bf \ddot n}^{(0)} = - \omega^2 {\bf K}\left({\boldsymbol \rho}^{(0)}_{\rm min}[{\bf n}^{(0)}] - {\bf n}^{(0)} \right),
\end{equation}
where ${\bf K} \in {\boldmath R}^{N \times N}$, ${\bf K} = {\bf J}^{-1}$, ${\boldsymbol \rho}^{(0)}_{\rm min}[{\bf n}^{(0)}] \in {\boldmath R}^N$, and ${\bf n}^{(0)} \in {\boldmath R}^N$.
We can rewrite this equation of motion in an equivalent preconditioned form,
\begin{equation}
{\bf \ddot n}^{(0)} = - \omega^2 \left({\bf K}_0{\bf J}\right)^{-1}{\bf K}_0\left({\boldsymbol \rho}^{(0)}_{\rm min}[{\bf n}^{(0)}] - {\bf n}^{(0)} \right),
\end{equation}
where we have introduced a preconditioner, ${\bf K}_0 \approx {\bf J}^{-1}$.
${\bf J}$ is the Jacobian of the residual function,
\begin{align}
{\bf f}({\bf n}^{(0)}) = & {\boldsymbol \rho}^{(0)}_{\rm min}[{\bf n}^{(0)}] - {\bf n}^{(0)}.
\end{align}
If we use the notation,
\begin{align}
{\bf f}_{{\bf v}_k}({\bf n}^{(0)}) \equiv & {\bf K}_0\frac{ \partial {\bf f}({\bf n}^{(0)} + \lambda {\bf v}_k)}{\partial \lambda} \Big \vert_{\lambda = 0} = {\bf K}_0{\bf J} {\bf v}_k,\label{DirDer}
\end{align}
it is possible to show that the preconditioned Jacobian, ${\bf K}_0{\bf J}$, can be
approximated by a low-rank (rank-$m$) approximation,
\begin{align}\label{rankm}
{\bf K}_0{\bf J} \approx \sum_{kl}^{m} {\bf f}_{{\bf v}_k} L_{kl} {\bf v}_l^{\rm T},
\end{align}
for some set of vectors $\{{\bf v}_k\}$, and with ${\bf L} = {\bf O}^{-1}$,
where $O_{ij} = {\bf v}_i^T{\bf v}_j$ and $m < N$ \cite{ANiklasson20}. The directional
derivatives of ${\bf f}({\bf n})$ in the direction of ${\bf v}_k$ (or Gateaux derivatives) in Eq.\ (\ref{DirDer}) 
can be calculated using quantum perturbation theory \cite{ANiklasson20,ANiklasson15,YNishimoto17}.
%The cost of each directional derivative is dominated by the construction of the Coulomb matrix and the exchange correlation matrix generated by the perturbing charge vector ${\bf v}_k$.
%This is also required in a regular SCF iteration, but no additional diagonalization is needed to calculated the directional derivatives.
%The cost of a directional derivative is thus like an SCF iteration, but without the additional and typically dominating cost of a diagonalization.

The low-rank inverse of the preconditioned Jacobian, ${\bf K}_0{\bf J}$, is then given by a pseudoinverse,
\begin{align}
\left({\bf K}_0{\bf J}\right)^{-1} \approx \sum_{kl}^m {\bf v}_k { M_{kl}} {{\bf f}^T_{{\bf v}_l}}, \label{PKSAP}
\end{align}
with ${\bf M} = {\bf O}^{-1}$, where $O_{ij} = {\bf f}_{{\bf v}_i}^T{\bf f}_{{\bf v}_j}$.
By chosing the vectors, $\{{\bf v}_k\}$, from an orthogonalized
preconditioned Krylov subspace \cite{ANiklasson20},
\begin{align}
& \left\{{\bf v}_k\right\} \in {\rm span}^\perp \left\{ {\bf K}_0{{\bf f}}({\bf n}^{(0)}), 
({\bf K}_0 {\bf J})^{1}{\bf K}_0{{\bf f}}({\bf n}^{(0)}), \right. \\ 
& \left. ({\bf K}_0 {\bf J})^2 {\bf K}_0{{\bf f}}({\bf n}^{(0)}),
({\bf K}_0 {\bf J})^3 {\bf K}_0{{\bf f}}({\bf n}^{(0)}), \ldots  \right\},
\end{align}
we can rapidly reach a well-converged approximation of the kernel, ${\bf K}$, acting
on the residual function.  The advantage with the preconditioner, ${\bf K}_0$, is that it typically reduces the number of Krylov subspace vectors (or low-rank updates) necessary to reach convergence. However, in principle the preconditioner is not needed and the low-rank Krylov subspace approximation works well also without preconditioning \cite{VGavini22}.
%Generating each new vector carries the approximate cost of an SCF iteration, but without a diagonalization.
%For the
%integration of the electronic equation of motion, Eq.\ (\ref{EOM_n}), we then get
%\begin{align}
%{\bf \ddot n}^{(0)} = - \omega^2 \left(\sum_{kl} {\bf v}_k { M_{kl}} {{\bf f}_{{\bf v}_l}} \right){\bf K}_0\left({\boldsymbol \rho}^{(0)}_{\rm min}[{\bf n}^{(0)}] - {\bf n}^{(0)} \right),
%\end{align}
%and for the Newton step in Eq.\ (\ref{Newton}) we get
%\begin{equation}\label{Newton2}\begin{array}{l}
%{\bf n}^{(1)} = {\bf n}^{(0)} - \left(\sum_{kl} {\bf v}_k {\widetilde M_{kl}} {\widetilde {\bf f}_{{\bf v}_l}} \right){\bf K}_0\left({\boldsymbol \rho}^{(0)}_{\rm min}[{\bf n}^{(0)}] - {\bf n}^{(0)} \right).
%\end{array}
%\end{equation}

If we let $\Delta {\bf n}^{(0)}$ denote the result of the kernel acting on the residual, i.e.\
\begin{align}\label{KRes}
%{\bf K} {\bf f}[{\bf n}^{(0)}]  \approx  - \left(\sum_{kl} {\bf v}_k {\widetilde M_{kl}} {\widetilde {\bf f}_{{\bf v}_l}} \right){\bf K}_0\left({\boldsymbol \rho}^{(0)}_{\rm min}[{\bf n}^{(0)}] - {\bf n}^{(0)} \right),
\Delta {\bf n}^{(0)}  = & \left({\bf K}_0{\bf J}\right)^{-1} {\bf K}_0 \left({\boldsymbol \rho}^{(0)}_{\rm min}[{\bf n}^{(0)}] - {\bf n}^{(0)} \right)\\
 %\approx & \left(\sum_{kl} {\bf v}_k {M_{kl}} {{\bf f}^T_{{\bf v}_l}} \right){\bf K}_0\left({\boldsymbol \rho}^{(0)}_{\rm min}[{\bf n}^{(0)}] - {\bf n}^{(0)} \right),
 \approx & \left(\sum_{kl} {\bf v}_k {M_{kl}} {{\bf f}^T_{{\bf v}_l}} \right){\bf K}_0{\bf f}({\bf n}^{(0)}),\label{Kres_v}
\end{align}
we find that the electronic equation of motion in Eq.\ (\ref{EOM_n})
and the Newton step in Eq.\ (\ref{Newton}) are given by
\begin{align}
&{\bf \ddot n}^{(0)} = -\omega^2 \Delta {\bf n}^{(0)}, \label{d2ndt2}\\
&{\bf n}^{(1)} = {\bf n}^{(0)}  - \Delta {\bf n}^{(0)} \label{dn_newton}.
\end{align}
This clearly shows how the approximation of $\Delta {\bf n}^{(0)}$ only needs to be performed once every time step for
the $1$st-level generalized shadow XL-BOMD in Eqs.\ (\ref{EOM_R}) and (\ref{EOM_n}). This simplification is another of our key results.

The cost of calculating a preconditioner, ${\bf K}_0$, can be expensive, but in QMD simulations the preconditioner can be reused, often over thousands of integration time steps (or the whole simulation) before an updated preconditioner is needed. In practice the overhead is therefore quite small. In simulations of regular stable molecular systems a scaled delta function typically works perfectly well as a preconditioner. The main cost of the preconditioned subspace expansion required to approximate $\Delta {\bf n}^{(0)}$ in Eq.\ (\ref{Kres_v}) is therefore the construction of the residual response vectors, $\{{\bf f}_{{\bf v}_k}\}$, from the directional perturbations in $\{{\bf v}_k\}$. In Kohn-Sham DFT these response vectors can be calculated from quantum perturbation theory 
\cite{ANiklasson15,YNishimoto17}.
If we assume that we have already performed a diagonalization of the unperturbed Kohn-Sham Hamiltonian, ${\bf H}[n^{(0)}]$, to find $\rho_{\rm min}[n^{(0)}]$ in Eq.\ (\ref{rho0_min}), these response vectors are fairly easy to calculate as no additional diagonalizations are needed \cite{VGavini22,CNegre22}. Nevertheless, the calculation of the residual response vectors, $\{{\bf f}_{{\bf v}_k}\}$, is the main bottleneck of the Krylov subspace expansion. In Kohn-Sham DFT, using an atomic-orbital basis, the cost is typically dominated by the transformations back and forth between the non-orthogonal atomic-orbital basis and the molecular-orbital eigenbasis, in which the response summations are performed \cite{ANiklasson20,ANiklasson20b}. 

%To construct each new vector 
%{\bf ABOUT THE COST}

\subsection{The Harris-Foulkes functional}

The $0$th and $1$st-level pairs of shadow energy functionals and Born-Oppenheimer potentials presented in this article, e.g.\ as in Eqs.\ (\ref{E0})-(\ref{U0}), are quite general and easy to apply in different applications, e.g.\ to orbital-based Kohn-Sham DFT, density matrix methods in Hartree-Fock theory, to orbital-free polarizable charge equilibration models, or in a slightly different form even to time-dependent models for superfluidity \cite{ANiklasson21b,PHenning21}. These linearized shadow energy functionals and their constrained optimization may, of course, appear somewhat trivial. It is only in combination with XL-BOMD that the shadow energy functionals and Born-Oppenheimer potentials become useful and consequential.
%It is only in combination with the extended Lagrangian dynamics that our shadow energy functionals and Born-Oppenheimer potentials become meaningful and consequential. 
The proposed $0$th or $1$st-level shadow energy functionals and potentials are only meaningful within the dynamical simulation framework of XL-BOMD, where the electronic degrees of freedom appear as dynamical field variables in addition to the nuclear positions and velocities. Only then can we calculate the exact forces from the shadow Born-Oppenheimer potential.
%It is therefore only in combination with the extended Lagrangian dynamics that our shadow energy functionals and Born-Oppenheimer potentials become useful and consequential. 
For a static, non-dynamical problem,
the corresponding forces are only approximate, and would, in general, require a well-converged iterative SCF optimization procedure to achieve any reasonable accuracy.

For the static, non-dynamical problem, and for the particular case of Kohn-Sham DFT, the optimized ground-state shadow Born-Oppenheimer potential, although conceptually different, is interchangeable with the Harris-Foulkes energy density functional  \cite{JHarris85,MFoulkes89, ANiklasson14}. 
The Harris-Foulkes (HF) functional, $ E_{\rm HF}[\rho_0]$, is an {\em approximate} energy expression for the electronic ground-state energy in Kohn-Sham DFT that depends on some input density, $\rho_0({\bf r})$, where
\begin{align}
    E_{\rm HF}[\rho_0] = & \sum_i f_i \varepsilon_i - \frac{1}{2} \iint \frac{\rho_0({\bf r}) \rho_0({\bf r'})}{\vert {\bf r-r'}\vert} d{\bf r}d{\bf r'}\\
    & + E_{\rm xc}[\rho_0] - \int V_{\rm xc}[\rho_0]({\bf r})\rho_0({\bf r}) d{\bf r}.
\end{align}
Here $\{\varepsilon_i\}$ are the eigenvalues of the Kohn-Sham Hamiltonian, $H_{\rm KS}[\rho_0]$, calculated for the input density, $\rho_0({\bf r})$, $\{f_i\}$ are the occupation numbers, $E_{\rm xc}[\rho_0]$ is the exchange-correlation energy functional with the corresponding exchange-correlation potential, $V_{\rm xc}[\rho_0]({\bf r}) = \delta E_{\rm xc}[\rho]\big / \delta \rho({\bf r})\big \vert_{\rho_0}$. 

Apart from the nuclear-nuclear repulsion term (and possibly an additional electronic entropy contribution), $E_{\rm HF}[\rho_0]$, has the same form as the $0$th-level shadow potential, ${\cal U}^{(0)}({\bf R},n^{(0)})$, with $\rho_0 = n^{(0)}$.
The difference is that ${\cal U}^{(0)}({\bf R},n^{(0)})$, as it appears in XL-BOMD, represents an {\em exact} ground-state shadow Born-Oppenheimer potential, which is determined from a variationally optimized shadow energy functional, ${\cal E}^{(0)}[\rho,n^{(0)}]$, with some external and electrostatic potentials that are given by the nuclear positions, ${\bf R}(t)$, and a separate {\em dynamical variable} density, $n^{(0)}({\bf r},t)$. Because $n^{(0)}({\bf r},t)$ is a dynamical field variable of the extended Lagrangian in Eq.\ (\ref{XL1}), forces in the Euler-Lagrange's equations of motion can easily be calculated from the partial derivatives of ${\cal U}^{(m)}({\bf R},n^{(0)})$ with respect to a constant density, $n^{(0)}({\bf r},t)$. 
%The shadow Born-Oppenheimer potential can also easily be designed for a broad variety of energy expressions beyond Kohn-Sham DFT, including Hartree-Fock theory and orbital-free flexible charge equilibration models \cite{ANiklasson21b}.
This is in contrast to the Harris-Foulkes functional, which is an approximate energy density functional expression for the (static) Kohn-Sham ground state energy, where the density $\rho_0({\bf r})$ represents, either overlapping ${\bf R}$-dependent atomic charge densities, or some iteratively and partially SCF updated (and thus ${\bf R}$-dependent) input density. The Harris-Foulkes energy functional is thus best used for estimating the electronic ground state energy for approximate densities. Accurate calculations of the interatomic forces would still require a regular iterative SCF optimization, or the additional calculation of the gradients of the electron density with respect to the atomic positions. 

In XL-BOMD the pairs of shadow energy functionals and potential energy surfaces therefore play a different role, and allows for computationally simple and accurate calculations of conservative interatomic forces in molecular dynamics simulations, without relying on the Hellmann-Feynman theorem \cite{DMarx00,RPFeynman39}. However, the linearized shadow energy functionals and optimized Born-Oppenheimer potentials presented and derived here, as in Eqs.\ (\ref{E0})-(\ref{U0}), provide an alternative and probably more transparent and straightforward approach to derive and understand the Harris-Foulkes functional in Kohn-Sham DFT. The procedure in Eqs.\ (\ref{E0})-(\ref{U0}) is also easy to generalize and apply to a broad variety of other energy expressions besides the Kohn-Sham energy functional \cite{ANiklasson21b}. As an example, in the section below, we will use the approach in Eqs.\ (\ref{E0})-(\ref{U0}) to the design a shadow energy functional and potential for a coarse-grained flexible charge equilibration model.

\subsection{Coarse-grained flexible charge model}

Flexible charge models can be derived from an atomic coarse-graining of DFT \cite{FJVesely77,MSprik88,WJMortier86,AKRappe91,GLamoureaux03,SNaserifar17,DMYork96,GTabacchi02,TVerstraelen13,ANiklasson21,ANiklasson21b}. They often serve as simplified or conceptual versions of DFT and can also be used to illustrate our shadow energy functionals in Born-Oppenheimer molecular dynamics.

In the simplest form of  flexible charge models the electronic energy functional in DFT is approximated by the energy function
\begin{align}
E({\bf R,q}) = \sum_I \chi_I q_I + \frac{1}{2} \sum_I U_I q_I^2 + \frac{1}{2} \sum_{IJ}^{I \ne J} q_I\gamma_{IJ}q_J,
\end{align}
where ${\bf q} = \{q_I\}$ is the coarse-grained charge density, represented by net partial charges (or electron occupations) of each atom $I$, $\chi_I$ are the estimated atomic electronegativities, $U_I$ the chemical hardness or Hubbard-U parameters, and $\gamma_{IJ}$ describe the Coulomb interactions between penetrating spherical atom-centered charge densities centered at atom $I$ and $J$. At large interatomic distances these interactions decay as $\gamma_{IJ} \rightarrow  \vert {\bf R}_I - {\bf R}_J \vert^{-1}$ and at short-range distances the onsite limit, $\gamma_{IJ} \rightarrow U_I$, is reached as $\vert {\bf R}_I - {\bf R}_J \vert \rightarrow 0$.

The electronic ground state is given from the constrained minimization, where
\begin{align}
{\bf q}_{\rm min} = \min_{\bf q} \left\{ E({\bf R,q}) \left \vert \sum_I q_I = 0 \right. \right\}.
\end{align}
This minimization requires the solution of a full system of linear equations, which is the main computational bottleneck. If an iterative solver is used the optimized solutions need to be well-converged to provide accurate conservative forces in a molecular dynamics simulation.
The optimized ground state charges then gives us the Born-Oppenheimer potential, 
\begin{align}
U({\bf R}) = E({\bf R},{\bf q}_{\rm min}) + V({\bf R}). 
\end{align}
The molecular trajectories can then be generated from the integration of Newton's equations of motion,
\begin{align}
    M_I {\bf \ddot R}_I = - \nabla_I U({\bf R}).
\end{align}

Following the approach in Eqs.\ (\ref{E0})-(\ref{U0}), a $0$th-level shadow energy function, ${\cal E}^{(0)}({\bf R,q},{\bf n}^{(0)}) \approx E({\bf R,q})$, can be constructed from a partial linearization of $E({\bf R,q})$ around some approximate ground state solution, ${\bf n}^{(0)} \approx {\bf q}_{\rm min}$, where
\begin{align}
{\cal E}^{(0)}({\bf R,q},{\bf n}^{(0)}) =& \sum_I \chi_i q_i + \frac{1}{2} \sum_I U_Iq_I^2 \\
&+ \frac{1}{2} \sum_{I \ne J} (2q_I-n_I^{(0)})\gamma_{IJ}n_J^{(0)}.
\end{align}
The constrained minimization (the lowest stationary solution) of this shadow energy function gives us the ${\bf n}^{(0)}$-dependent ground state density,
\begin{align}
{\bf q}_{\rm min}[{\bf n}^{(0)}] = \arg \min_{\bf q} \left\{ {\cal E}^{(0)}({\bf R,q},{\bf n}^{(0)}) \left \vert \sum_I q_I = 0 \right. \right\}
\end{align}
and the corresponding $0$th-level shadow Born-Oppenheimer potential,
\begin{align}
{\cal U}^{(0)}({\bf R},{\bf n}^{(0)}) = {\cal E}^{(0)}({\bf R},{\bf q}_{\rm min}[{\bf n}^{(0)}], {\bf n}^{(0)}) + V({\bf R}).
\end{align}
The shadow energy function, ${\cal E}^{(0)}({\bf R,q},{\bf n}^{(0)})$, is constructed such that ${\bf q}_{\rm min}[{\bf n}^{(0)}]$ is determined by a quasi-diagonal system of linear equations that has a trivial analytical solution \cite{ANiklasson21,ANiklasson21b}. 

To introduce the 1st-level update we can improve the ground-state estimate of ${\bf n}^{(0)}$ with a Newton step, 
\begin{align}
 &{\bf n}^{(1)} \equiv {\bf n}^{(1)}[{\bf n}^{(0)}]= {\bf n}^{(0)} - \Delta {\bf n}^{(0)},
\end{align}
where
\begin{align}
\Delta {\bf n}^{(0)}  = & \left({\bf K}_0{\bf J}\right)^{-1} {\bf K}_0 \left({\boldsymbol \rho}^{(0)}_{\rm min}[{\bf n}^{(0)}] - {\bf n}^{(0)} \right),
\end{align}
which can be approximated, for example, by the preconditioned low-rank Newton step as in Eq.\ (\ref{Kres_v}).
Notice that this updated approximate charge vector is ${\bf n}^{(0)}$-dependent, i.e.\
\begin{align}
 &{\bf n}^{(1)} \equiv {\bf n}^{(1)}[{\bf n}^{(0)}]= {\bf n}^{(0)} - \Delta {\bf n}^{(0)}.
\end{align}
The updated $1$st-level energy function is now given by
\begin{align}
{\cal E}^{(1)}({\bf R,q},{\bf n}^{(1)}) =& \sum_I \chi_i q_i + \frac{1}{2} \sum_I U_I q_I^2 \\
&+ \frac{1}{2} \sum_{I \ne J} (2q_I-n_I^{(1)})\gamma_{IJ}n_J^{(1)}.
\end{align}
The optimized ground state density is then given from the constrained minimization, where
\begin{align}
{\bf q}_{\rm min}[{\bf n}^{(1)}] 
 = \arg \min_{\bf q} \left\{ {\cal E}^{(1)}({\bf R,q},{\bf n}^{(1)}) \left \vert \sum_I q_I = 0 \right. \right\}.
\end{align}

The shadow energy function ${\cal E}^{(1)}({\bf R,q},{\bf n}^{(1)})$ is constructed in the same way as ${\cal E}^{(0)}({\bf R,q},{\bf n}^{(0)})$ such that ${\bf q}_{\rm min}[{\bf n}^{(1)}]$ also is determined by a quasi-diagonal system of linear equations that has a trivial analytical solution \cite{ANiklasson21,ANiklasson21b}.
This gives us the corresponding $1$st-level shadow Born-Oppenheimer potential,
\begin{align}
{\cal U}^{(1)}({\bf R},{\bf n}^{(0)}) = {\cal E}^{(1)}({\bf R},{\bf q}_{\rm min}[{\bf n}^{(1)}],{\bf n}^{(1)}) + V({\bf R}),
\end{align}
where ${\bf n}^{(1)} \equiv {\bf n}^{(1)}[{\bf n}^{(0)}]$.
The shadow Born-Oppenheimer potential can then be used in an extended Lagrangian formulation \cite{ANiklasson21,ANiklasson21b}, which in an adiabatic limit gives us the equations of motion,
\begin{align}
    &M_I {\bf \ddot R}_I = - \nabla_I {\cal U}^{(1)}({\bf R},{\bf n}^{(0)})\big \vert_{{\bf n}^{(0)}},\\
     &{\bf \ddot n}^{(0)} = - \omega^2 \Delta {\bf n}^{(0)}.
\end{align}
The nuclear coordinates and velocities can then be integrated using a standard velocity Verlet integration scheme and for the evolution of the atomic partial charges, ${\bf n}(t)$, we can use the modified Verlet integation scheme including some additional weak dissipative damping forces as in Eq.\ (\ref{Int_n}).

This example with a coarse-grained flexible charge equilibration model demonstrates the general applicability of our shadow molecular dynamics approach and how it can be used to construct pairs of shadow energy functionals and potentials for XL-BOMD simulations at different levels of accuracy.

\section{Pseudocode}

The easiest way to summarize the generalized $1$st-level update of the shadow energy functional and Born-Oppenheimer potential in XL-BOMD is to describe the method in a step-by-step procedure using a pseudocode.
Algorithm \ref{XLBOMD_SCHEME} gives a schematic picture of what an XL-BOMD simulation using the $1$st-level shadow potential, ${\cal U}^{(1)}$, would look like for an orbital-dependent Kohn-Sham like electronic structure theory. It is expressed in a matrix-vector notation that is well-suited, for example, for SCC-DFTB simulations. All $0$th-level superscript, $^{(0)}$, as in ${\bf n}^{(0)}$, have been dropped to simplify the notation.
Here ${\bf S}$ is a basis-set overlap matrix and ${\bf H}$ is the effective single-particle (Kohn-Sham) Hamiltonian. Of critical importance is the construction of $\Delta {\bf n} \equiv \Delta {\bf n}^{(0)} $ with a low-rank preconditioned Krylov subspace approximation
using quantum response calculations. In contrast to the most recent XL-BOMD schemes, we now need two diagonalizations per time step,
instead of only one. Algorithm \ref{XLBOMD_SCHEME} provides a compact summary of the most important results of this article.

\begin{algorithm}[H]
\caption{Pseudocode for the XL-BOMD scheme using the $1$st-level updated
shadow Born-Oppenheimer potential, ${\cal U}^{(1)}({\bf R},n^{(0)})$.
Matrix-vector notation is used and the $0$th-level ${(0)}$-superscripts, i.e.\ as in ${\bf n}^{(0)}$
or $\Delta {\bf n}^{(0)}$, have been dropped for brevity. One rank-$m$ approximation of $\Delta {\bf n}$
and two Hamiltonian diagonalizations are required in each time step.}\label{XLBOMD_SCHEME}
\algsetup{indent=1em}
\begin{algorithmic}
\STATE $ \mbox{Atomic masses and positions,~} {\bf M} = \{M_I\},~{\bf R} = \{{\bf R}_I\}$
\STATE $ \mbox{Get ground state,~}{\bf q}_{\rm min}, \mbox{~with regular SCF}$
\STATE $ {\bf q}_{\rm min} \Rightarrow \mbox{``exact''}~{U}({\bf R})~\mbox{and forces}, {\bf F} = \{{\bf F}_I\}$
\STATE $ \mbox{Initialize charges,}~{\bf n}_j = {\bf q}_{\rm min},~ j = 1,2,\ldots,k $
\STATE $ \mbox{Initialize velocities,}~{\bf V} = \{{\bf V}_I\}$
\STATE $ \mbox{Estimate preconditioner,}~ {\bf K}_0 = {\bf J}^{-1}$
\STATE $ \mbox{Initial}~\Delta {\bf n} = ({\bf K}_0{\bf J})^{-1}{\bf K}_0({\bf q}_{\rm min}[{\bf n}] -{\bf n}) = {\bf 0}$
\STATE $ t = t_0$
\WHILE{$t < t_{\rm max}$}
 \STATE ${\bf V}_I = {\bf V}_I + (\delta t/2){\bf F}_I/M_I$
 \STATE ${\bf n}_0 = 2{\bf n}_1 - {\bf n}_2 - \delta t^2 \omega^2 \Delta {\bf n} + \alpha \sum_{l=0}^{k} c_l {\bf n}_{1-l}$
 %\mbox{Damp}({\bf n}_1,{\bf n}_2, \ldots, {\bf n}_k)$
 \STATE ${\bf n}_k = {\bf n}_{k-1},~ \ldots,~ {\bf n}_2 = {\bf n}_1,~  {\bf n}_1 = {\bf n}_0, ~ {\bf n} = {\bf n}_0$
 \STATE ${\bf R}_I = {\bf R}_I + \delta t{\bf V}_I$
% \STATE ${\bf S} = {\bf S}[{\bf R}],~ {\bf Z} = {\bf S}^{-1/2}$
 \STATE ${\bf H}[{\bf n}] = {\bf H}[{\bf R},{\bf n}],~{\bf S} = {\bf S}[{\bf R}],~ {\bf Z} = {\bf S}^{-1/2}$
 \STATE ${\bf q}_{\rm min}[{\bf n}] \Leftarrow \mbox{from diagonalized}~{\bf Z}^T{\bf H}[{\bf n}]{\bf Z}$
 \STATE $ \Delta {\bf n} = ({\bf K}_0{\bf J})^{-1}{\bf K}_0({\bf q}_{\rm min}[{\bf n}] -{\bf n}) ~\mbox{with~rank-}m ~\mbox{approx.}$
 \STATE $ {\bf n}^{(1)} = {\bf n} - \Delta {\bf n}, ~ \mbox{approximate Newton step}$ 
 \STATE ${\bf q}_{\rm min}[{\bf n}^{(1)}] \Leftarrow \mbox{from diagonalized}~{\bf Z}^T{\bf H}[{\bf n}^{(1)}]{\bf Z}$
 \STATE $ {\bf q}_{\rm min}[{\bf n}^{(1)}] \Rightarrow \mbox{shadow} ~{\cal U}^{(1)}({\bf R},{\bf n})~\mbox{and forces},~ {\bf F}$
 \STATE ${\bf V}_I  = {\bf V}_I + (\delta t/2){\bf F}_I/M_I$
 \STATE $ t = t + \delta t$
\ENDWHILE
\end{algorithmic}
\end{algorithm} 

%only has to be calculated once in our XL-BOMD scheme using the $1$st-level generalized schadow Born-Oppenheimer
%potential as in Eqs.\ (\ref{EOM_R}) and (\ref{EOM_n}).
%Using this $\Delta {\bf n}^{(0)}$-notation above, we thus get
%\begin{align}
%&{\bf \ddot n}^{(0)} = \omega^2 \Delta {\bf n}^{(0)},\\
%&{\bf n}^{(1)} = {\bf n}^{(0)}  + \Delta {\bf n}^{(0)}.
%\end{align}

%{\bf How the Krylov subspace is built using quantum-perturbation theory with the known eigenvectors has to be 
%mentioned as well}

\section{Examples}

We will demonstrate the accuracy and performance of the shadow energy functionals and Born-Oppenheimer potentials in XL-BOMD simulations using SCC-DFTB theory \cite{WHarrison80,MFoulkes89,DPorezag95,MElstner98,MFinnis98,TFrauenheim00,PKoskinen09,MGaus11,BAradi15,BHourahine20}.
SCC-DFTB theory can be seen as a framework for different levels of approximations of density functional theory. 
Here we will use the scheme given by a second-order expansion in the charge density fluctuations
around a reference density of overlapping neutral atomic charge distributions,
where the atomic net Mulliken partial charges are used to describe the long-range electrostatic interactions.
In this way the continuous charge density, $\rho({\bf r})$, of regular DFT becomes vectorized with one net partial charge per atom, ${\bf q} = \{q_I\}$.
The fluctuating partial charges are optimized self-consistently to account for interatomic charge transfer and the response to the long-range electrostatic interactions. In a general SCC-DFTB scheme this requires a repeated set of constructions of an approximate effective single-particle Kohn-Sham Hamiltonian, diagonalizations, charge calculations from the eigenfunctions, and Coulomb potential summations, until a self-consistent charge convergence is reached.
SCC-DFTB theory therefore follows the same iterative SCF procedure 
as a regular first-principles Kohn-Sham DFT calculation. Here we will also use a thermal DFTB theory, where we assume fractional occupation numbers of the molecular orbitals determined by the Fermi function at some given electronic temperature, $T_e$, including an electronic entropy term \cite{NMermin63,NMermin65,RParr89,EEngel11,SPittalis11,APJones14,ANiklasson21b}.
The fractional occupation numbers are important to better stabilize the electronic
structure calculations when the electronic HOMO-LUMO energy gap is small or vanishing. This also affects how we perform the response calculations of $\{{\bf f}_{{\bf v}_i}\}$ in the Krylov subspace approximation in Eq.\ (\ref{PKSAP}) of the preconditioned kernel \cite{ANiklasson15,YNishimoto17,ANiklasson17,ANiklasson21b,VGavini22,CNegre22}.

For our implementation and XL-BOMD simulations we use a developers version of the LATTE software package \cite{LATTE,MCawkwell12,AKrishnapriyan17} that closely follows Alg.\ \ref{XLBOMD_SCHEME}.
As preconditioner we use an exact calculation of the kernel in the first time step, and we use a sufficient number of low-rank updates to achieve an approximate quadratic convergence in the Newton updates. The maximum number, $m$, of Krylov subspace vectors, i.e.\ in the rank-$m$ approximation, never exceeds 6.

First we will look at the asserted scaling expressed in Eqs.\ (\ref{Res0}) and (\ref{Res1}) that were assumed 
in the derivation of the equations of motion, in Eq.\ (\ref{EOM_R}) and Eq.\ (\ref{EOM_n}).
Thereafter, we will demonstrate the advantage of the $1$st-level update of the shadow energy functional and Born-Oppenheimer potential, ${\cal U}^{(1)}$, compared to the original $0$th-level approximation for XL-BOMD simulations of an unstable, charge-sensitive, chemical system.

\begin{figure}
\includegraphics[scale=0.33]{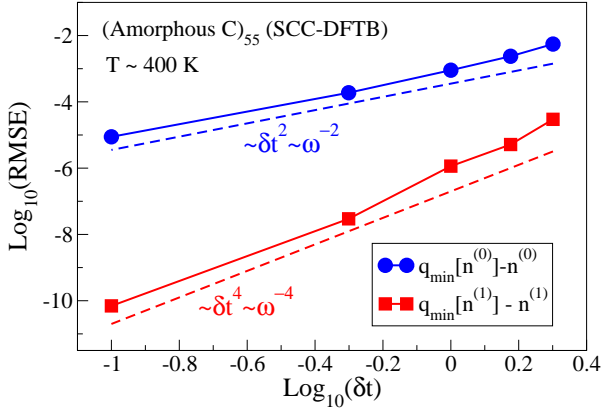}
\caption{\label{Fig_1}
{\small Scaling of the residual error terms as a function of time step, $\delta t$,
or harmonic oscillator frequency, $\omega$, for a system of amorphous carbon with 55 atoms using periodic boundary conditions.
The simulations are performed with a constant dimensionless constant, $\kappa = \delta t^2 \omega^2$, which means that $\delta t \propto \omega^{-1}$.
XL-BOMD based on the enhanced $1$st-level shadow Born-Oppenheimer potential, ${\cal U}^{(1)}({\bf R},n^{(0)})$, was used following Alg.\ \ref{XLBOMD_SCHEME}.
The root-mean square errors (RMSE) are given by the root-mean-square of the residuals, ${\bf q}_{\rm min}[{\bf n}^{(0)}]-{\bf n}^{(0)}$
and $q_{\rm min}[{\bf n}^{(1)}]-{\bf n}^{(1)}$, and are averaged over snapshots of 100 integration time steps.
The dashed lines indicates the exact $\delta t^2 \sim \omega^{-2}$ and $\delta t^4 \sim \omega^{-4}$ scalings.}}
\end{figure}

\subsection{Scaling}

Figure \ref{Fig_1} shows the approximate scaling of the root mean square errors (RMSE) given by the root mean square of the residuals, $ {\bf q}_{\rm min}[{\bf n}^{(0)}] - {\bf n}^{(0)} $
%\propto \omega^{-2}$
and $ {\bf q}_{\rm min}[{\bf n}^{(1)}] - {\bf n}^{(1)}$
%\vert\propto \omega^{-4}$, 
for simulations of amorphous carbon. 
The results of the simulations confirm the assumed scaling orders, where $\vert {\bf q}_{\rm min}[{\bf n}^{(0)}] - {\bf n}^{(0)} \vert \propto \omega^{-2}$
and $\vert {\bf q}_{\rm min}[{\bf n}^{(1)}] - {\bf n}^{(1)} \vert\propto \omega^{-4}$.
These scalings were asserted {\em a priori} in the derivation of the equations of motion 
in an adiabatic limit as $\omega \rightarrow \infty$.
The scaling of the RMSE extracted from the XL-BOMD simulations shown in Fig.\ \ref{Fig_1} confirms these assumption.
Notice that the $\omega^{-1} \propto \delta t$, because our integration scheme, Eq.\ (\ref{Int_n}), has been chosen such that $\delta t^2 \omega^2$ is a dimensionless constant, ${\kappa} = \delta t^2 \omega^2$.

\begin{figure}
\includegraphics[scale=0.33]{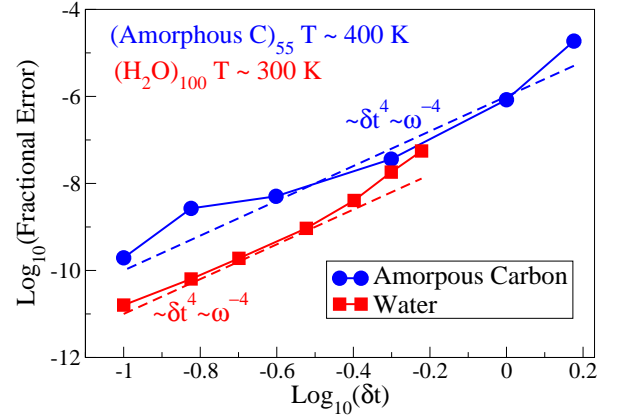}
\caption{\label{Fig_2}
{\small Scaling of the fractional error in the interatomic forces for water and amorphous carbon
as a function of the integation time step $\delta t$, or harmonic oscillator frequency, $\omega$.
XL-BOMD based on the enhanced $1$st-level shadow potential, ${\cal U}^{(1)}({\bf R},n^{(0)})$, was used.
The simulations are performed with a constant dimensionless constant, $\kappa = \delta t^2 \omega^2$, which means that $\delta t \propto \omega^{-1}$.
The fractional error was estimated from an on-the-fly comparison with the ``exact'' fully converged
Born-Oppenheimer forces, where the error was averaged over all the atoms and force components 
over a snapshot of 100 integration time steps. The dashed lines indicates the exact $\sim \delta t^4$ or $\sim \omega^{-4}$ scalings.}}
\end{figure}

The error in the $0$th-level shadow Born-Oppenheimer potential scales as $\vert {\cal U}^{(0)} - U\vert \propto \vert {\bf q}_{\rm min}[{\bf n}^{(0)}] - {\bf n}^{(0)} \vert^2$.
This means that the error in the sampling of the $0$th-level shadow Born-Oppenheimer potential, ${\cal U}^{(0})$, scales as $\delta t^4$ with the integration time step, which has been confirmed previously, e.g.\ \cite{ANiklasson17}.
The new $1$st-level updated shadow Born-Oppenheimer potential, ${\cal U}^{(1)}$, has the same form for the error, where
$\vert {\cal U}^{(1)} - U\vert \propto \vert {\bf q}_{\rm min}[{\bf n}^{(1)}] - {\bf n}^{(1)} \vert^2$.
This means, from the scaling demonstrated in Fig.\ \ref{Fig_1}, that the error in the sampling of the shadow Born-Oppenheimer potential ${\cal U}^{(1)}$ scales at $\delta t^8$. It is hard to demonstrate this scaling of the error in ${\cal U}^{(1)}$ directly, because the error converges to quickly and saturates at a level set by the available numerical precision. Here we therefore only show this $\delta t^8$-scaling indirectly, from the $\delta t^4$ or $\omega^{-4}$-scaling of $\vert {\bf q}_{\rm min}[{\bf n}^{(1)}] - {\bf n}^{(1)}\vert$ in Fig.\ \ref{Fig_1}. 

From the derivation of the equations of motion with the $1$st-level shadow potential in Eqs.\ (\ref{EOM_R}) and (\ref{EOM_n}), we made the estimate that the equations of motion 
for the atomic positions should have an error that scales as $\propto \omega^{-4}$. In Fig.\ \ref{Fig_2}
we show the results of simulations of an amorphous Carbon and a water system, were we find that the fractional error in the evaluated forces for the $1$st-level ${\cal U}^{(1)}$ shadow potential scale at $\propto \delta t^4$. This confirms the previously estimated scaling. 
This is in contrast to the original $0$th-level shadow Hamiltonian formulation of XL-BOMD using ${\cal U}^{(0)}$ with an error in the forces that is only of second order, $\propto \delta t^2$ \cite{ANiklasson17}.

The dramatic improvement in the scaling of the error as a function of the integration
time step may seem impressive. Nevertheless, often the improved behavior only has a minor effect on the accuracy and stability of XL-BOMD simulations. 
It is only for highly unstable systems, where the improved scaling and accuracy from the $1th$-level update of the shadow energy functional and Born-Oppenheimer potential 
play a role. For such problems we find that stable molecular trajectories
often can be achieved with a slightly longer integration time step than what otherwise
would be possible with the original $0$th-level shadow energy functional and Born-Oppenheimer potential.

Another important observation is that the higher-degree of accuracy in the force evaluations may be useful if higher-order symplectic integration schemes are used. In previous studies, using earlier versions of XL-BOMD, we found that we needed a fairly tight SCF convergence prior to the force evaluations for the higher-order symplectic integration schemes in order to take full advantage of their improved accuracy \cite{AOdell09,AOdell11}. The $1$st-level shadow energy functional and Born-Oppenheimer potential should therefore be well-suited in combination with various 4th-order symplectic integrations schemes \cite{AOdell09}. 

\begin{figure}
\includegraphics[scale=0.31]{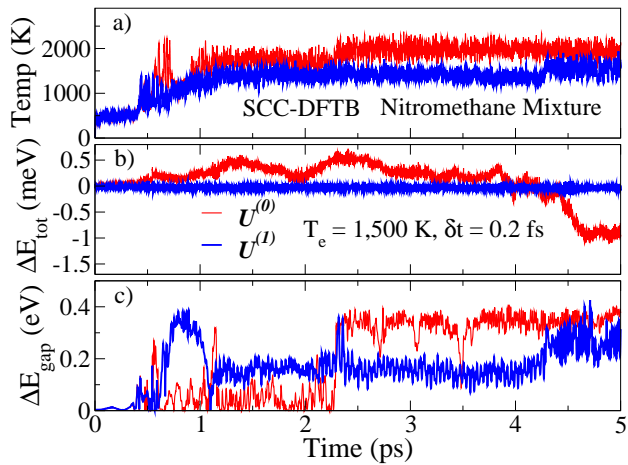}
\caption{\label{Fig_3}
{\small XL-BOMD simulations based on SCC-DFTB theory of an artificial highly reactive randomized mixture of liquid nitromethane (49 atoms with periodic boundary conditions). 
The upper panel a) shows the statistical temperature, the middle panel b) shows the fluctuations in the total energy per atom, and the lower panel c) shows the HOMO-LUMO electronic energy gap. An integration time step of $\delta t = 0.2$ fs was used in combination with a fractional occupation number corresponding to an electronic temperature, $T_e = 1,500$ K. The $1$st-level updated shadow potential, ${\cal U}^{(1)}$, (blue lines) shows a more stable dynamics without the more pronounced fluctuations in the total energy fluctuations of the $0$th-level shadow potential, ${\cal U}^{(0)}$, (red lines).}}
\end{figure}

\subsection{Unstable mixture of nitromethane}

To demonstrate the advantage of the $1$st-level shadow energy functional and Born-Oppenheimer potential compared to the original $0$th-level approach, we will look at a chemically unstable system, with a small or vanishing HOMO-LUMO energy gap. Such systems are often difficult to study, in particular with regular direct quantum-mechanical Born-Oppenheimer molecular dynamics methods.  
As an example we have chosen an artificial mixture of liquid nitromethane, (CH$_3$NO$_2$)$_7$, where a handful randomly chosen atoms have switched places. This artificial testbed system is highly unstable and exothermic reactions occurs within a few hundred femtoseconds. This is illustrated in Fig.\ \ref{Fig_3}. We find a significantly improved stability in the simulation with the $1$st-level updated shadow potential, ${\cal U}^{(1)}$, (blue solid lines) compared to the original $0$th-level shadow potential, ${\cal U}^{(0)}$, (red dashed lines) as indicated by the fluctuations in the total energy shown in the mid panel b).
Only by reducing the integration time step, $\delta t$, or possibly by increasing the
electronic temperature, is it possible to stabilize the XL-BOMD simulation using the original $0$th-level shadow potential.

\section{Summary and discussion}

In this article we have introduced a generalization of the shadow energy functionals and Born-Oppenheimer potentials
used in XL-BOMD. The original 0th-level shadow energy functional generates a Born-Oppenheimer potential that has an error in the fourth-order, ${\cal O}(\delta t^4)$, of the integration time step, $\delta t$, and with an error in the interatomic forces that is of second-order, ${\cal O}(\delta t^2)$.
With the 1st-level update the error in the potential energy can be reduced to scale as ${\cal O}(\delta t^8)$, where
the error in the calculated interatomic forces scales as ${\cal O}(\delta t^4)$. The main additional cost using the $1$st-level instead of the $0$th-level shadow potential is the cost of an extra Hamiltonian diagonalization.
We showed how this improved level of accuracy helps stabilize the integration of the molecular trajectories, which can be of particular importance for unstable, charge-sensitive, reactive systems with a small or vanishing electronic HOMO-LUMO energy gap. The improved scaling in the error of the potential and forces may also be of interest in the application of higher-order symplectic integration schemes \cite{ANiklasson08,AOdell09,AOdell11}. These higher-order schemes are of no use unless they can be matched by force evaluations with a comparable or higher level of accuracy. 

The ability to systematically improve the accuracy of the Born-Oppenheimer potential has many
similarities with earlier versions of XL-BOMD \cite{ANiklasson06,ANiklasson08,PSteneteg10}, where often a few SCF steps were needed prior to each force evaluation. However, with the detailed analysis supported by the concept of a shadow dynamics or a backward error analysis, we now have a more transparent description of why and when this is the case and how we can optimize the efficiency of our XL-BOMD simulations. The key idea is the construction of pairs of shadow energy functionals and potentials, where the shadow potential is given from an exact, yet computationally cheap, ground-state optimization of a linearized shadow energy functional. In combination with XL-BOMD, where the electronic degrees of freedom is propagated dynamically, the shadow Born-Oppenheimer potentials can then be used to calculate conservative interatomic forces that generates accurate molecular trajectories with long-term energy stability. 

The generalized shadow energy functionals and Born-Oppenheimer potentials were demonstrated using Kohn-Sham based SCC-DFTB theory. However, the underlying theory was derived in a general form that also applies to other electronic structure theories, inlcuding Hartee-Fock and orbital-free DFT.
As an example we also discussed an extension to flexible charge equilibration models, which can be derived as coarse-grained versions of Hohenberg-Kohn DFT. The higher-level generalization of the shadow energy functionals and Born-Oppenheimer potentials presented here are therefore applicable to a broad variety of electronic structure methods and flexible charge models within the framework of XL-BOMD.

\section{Acknowledgements}

This work is supported by the U.S. Department of Energy Office of Basic Energy Sciences (FWP LANLE8AN,``Next generation quantum-based molecular dynamic'')
and by the U.S. Department of Energy through the Los Alamos National Laboratory. Discussions with Joshua Finkelstein are gratefully acknowledged.
Los Alamos National Laboratory is operated by Triad National Security, LLC, for the National Nuclear Security
Administration of the U.S. Department of Energy Contract No. 892333218NCA000001.

\bibliography{mndo_new_xy}

\begin{thebibliography}{91}%
\makeatletter
\providecommand \@ifxundefined [1]{%
 \@ifx{#1\undefined}
}%
\providecommand \@ifnum [1]{%
 \ifnum #1\expandafter \@firstoftwo
 \else \expandafter \@secondoftwo
 \fi
}%
\providecommand \@ifx [1]{%
 \ifx #1\expandafter \@firstoftwo
 \else \expandafter \@secondoftwo
 \fi
}%
\providecommand \natexlab [1]{#1}%
\providecommand \enquote  [1]{``#1''}%
\providecommand \bibnamefont  [1]{#1}%
\providecommand \bibfnamefont [1]{#1}%
\providecommand \citenamefont [1]{#1}%
\providecommand \href@noop [0]{\@secondoftwo}%
\providecommand \href [0]{\begingroup \@sanitize@url \@href}%
\providecommand \@href[1]{\@@startlink{#1}\@@href}%
\providecommand \@@href[1]{\endgroup#1\@@endlink}%
\providecommand \@sanitize@url [0]{\catcode `\\12\catcode `\$12\catcode
  `\&12\catcode `\#12\catcode `\^12\catcode `\_12\catcode `\%12\relax}%
\providecommand \@@startlink[1]{}%
\providecommand \@@endlink[0]{}%
\providecommand \url  [0]{\begingroup\@sanitize@url \@url }%
\providecommand \@url [1]{\endgroup\@href {#1}{\urlprefix }}%
\providecommand \urlprefix  [0]{URL }%
\providecommand \Eprint [0]{\href }%
\providecommand \doibase [0]{http://dx.doi.org/}%
\providecommand \selectlanguage [0]{\@gobble}%
\providecommand \bibinfo  [0]{\@secondoftwo}%
\providecommand \bibfield  [0]{\@secondoftwo}%
\providecommand \translation [1]{[#1]}%
\providecommand \BibitemOpen [0]{}%
\providecommand \bibitemStop [0]{}%
\providecommand \bibitemNoStop [0]{.\EOS\space}%
\providecommand \EOS [0]{\spacefactor3000\relax}%
\providecommand \BibitemShut  [1]{\csname bibitem#1\endcsname}%
\let\auto@bib@innerbib\@empty
%</preamble>
\bibitem [{\citenamefont {Yoshida}(1990)}]{HYoshida90}%
  \BibitemOpen
  \bibfield  {author} {\bibinfo {author} {\bibfnamefont {H.}~\bibnamefont
  {Yoshida}},\ }\href {\doibase 10.1016/0375-9601(90)90092-3} {\bibfield
  {journal} {\bibinfo  {journal} {Phys. Lett. A}\ }\textbf {\bibinfo {volume}
  {150}},\ \bibinfo {pages} {262} (\bibinfo {year} {1990})}\BibitemShut
  {NoStop}%
\bibitem [{\citenamefont {Grebogi}\ \emph {et~al.}(1990)\citenamefont
  {Grebogi}, \citenamefont {Hammel}, \citenamefont {Yorke},\ and\ \citenamefont
  {Saur}}]{CGrebogi90}%
  \BibitemOpen
  \bibfield  {author} {\bibinfo {author} {\bibfnamefont {C.}~\bibnamefont
  {Grebogi}}, \bibinfo {author} {\bibfnamefont {S.~M.}\ \bibnamefont {Hammel}},
  \bibinfo {author} {\bibfnamefont {J.~A.}\ \bibnamefont {Yorke}}, \ and\
  \bibinfo {author} {\bibfnamefont {T.}~\bibnamefont {Saur}},\ }\href {\doibase
  10.1103/PhysRevLett.65.1527} {\bibfield  {journal} {\bibinfo  {journal}
  {Phys. Rev. Lett.}\ }\textbf {\bibinfo {volume} {65}},\ \bibinfo {pages}
  {1527} (\bibinfo {year} {1990})}\BibitemShut {NoStop}%
\bibitem [{\citenamefont {Toxvaerd}(1994)}]{SToxvaerd94}%
  \BibitemOpen
  \bibfield  {author} {\bibinfo {author} {\bibfnamefont {S.}~\bibnamefont
  {Toxvaerd}},\ }\href {\doibase 10.1103/PhysRevE.50.2271} {\bibfield
  {journal} {\bibinfo  {journal} {Phys. Rev. E}\ }\textbf {\bibinfo {volume}
  {50}},\ \bibinfo {pages} {2271} (\bibinfo {year} {1994})}\BibitemShut
  {NoStop}%
\bibitem [{\citenamefont {Gans}\ and\ \citenamefont
  {Shalloway}(2000)}]{GJason00}%
  \BibitemOpen
  \bibfield  {author} {\bibinfo {author} {\bibfnamefont {J.}~\bibnamefont
  {Gans}}\ and\ \bibinfo {author} {\bibfnamefont {D.}~\bibnamefont
  {Shalloway}},\ }\href {\doibase 10.1103/PhysRevE.61.4587} {\bibfield
  {journal} {\bibinfo  {journal} {Phys. Rev. E}\ }\textbf {\bibinfo {volume}
  {61}},\ \bibinfo {pages} {4587} (\bibinfo {year} {2000})}\BibitemShut
  {NoStop}%
\bibitem [{\citenamefont {Bond}\ and\ \citenamefont
  {Leimkuhler}(2007)}]{ShadowHamiltonian}%
  \BibitemOpen
  \bibfield  {author} {\bibinfo {author} {\bibfnamefont {S.~D.}\ \bibnamefont
  {Bond}}\ and\ \bibinfo {author} {\bibfnamefont {B.~J.}\ \bibnamefont
  {Leimkuhler}},\ }\href@noop {} {\emph {\bibinfo {title} {Molecular dynamics
  and the accuracy of numerically computed averages}}}\ (\bibinfo  {publisher}
  {Cambride University Press},\ \bibinfo {address} {United Kingdom},\ \bibinfo
  {year} {2007})\BibitemShut {NoStop}%
\bibitem [{\citenamefont {Toxvaerd}\ \emph {et~al.}(2012)\citenamefont
  {Toxvaerd}, \citenamefont {Heilmann},\ and\ \citenamefont
  {Dyre}}]{SToxvaerd12}%
  \BibitemOpen
  \bibfield  {author} {\bibinfo {author} {\bibfnamefont {S.}~\bibnamefont
  {Toxvaerd}}, \bibinfo {author} {\bibfnamefont {O.~J.}\ \bibnamefont
  {Heilmann}}, \ and\ \bibinfo {author} {\bibfnamefont {J.~C.}\ \bibnamefont
  {Dyre}},\ }\href {\doibase 10.1063/1.4726728} {\bibfield  {journal} {\bibinfo
   {journal} {J. Chem. Phys.}\ }\textbf {\bibinfo {volume} {136}},\ \bibinfo
  {pages} {224106} (\bibinfo {year} {2012})}\BibitemShut {NoStop}%
\bibitem [{\citenamefont {Hammonds}\ and\ \citenamefont
  {Heyes}(2020)}]{KDHammonds20}%
  \BibitemOpen
  \bibfield  {author} {\bibinfo {author} {\bibfnamefont {K.~D.}\ \bibnamefont
  {Hammonds}}\ and\ \bibinfo {author} {\bibfnamefont {D.~M.}\ \bibnamefont
  {Heyes}},\ }\href {\doibase 10.1063/1.5139708} {\bibfield  {journal}
  {\bibinfo  {journal} {J. Chem. Phys.}\ }\textbf {\bibinfo {volume} {152}},\
  \bibinfo {pages} {024114} (\bibinfo {year} {2020})}\BibitemShut {NoStop}%
\bibitem [{\citenamefont {Hammonds}\ and\ \citenamefont
  {Heyes}(2021)}]{KDHammonds21}%
  \BibitemOpen
  \bibfield  {author} {\bibinfo {author} {\bibfnamefont {K.~D.}\ \bibnamefont
  {Hammonds}}\ and\ \bibinfo {author} {\bibfnamefont {D.~M.}\ \bibnamefont
  {Heyes}},\ }\href {\doibase 10.1063/5.0048194} {\bibfield  {journal}
  {\bibinfo  {journal} {J. Chem. Phys.}\ }\textbf {\bibinfo {volume} {154}},\
  \bibinfo {pages} {174102} (\bibinfo {year} {2021})}\BibitemShut {NoStop}%
\bibitem [{\citenamefont {Müser}(2022)}]{MHMuser22}%
  \BibitemOpen
  \bibfield  {author} {\bibinfo {author} {\bibfnamefont {M.~H.}\ \bibnamefont
  {Müser}},\ }\href {\doibase 10.1080/08927022.2022.2094430} {\bibfield
  {journal} {\bibinfo  {journal} {Molecular Simulation}\ }\textbf {\bibinfo
  {volume} {48}},\ \bibinfo {pages} {1393} (\bibinfo {year} {2022})},\ \Eprint
  {http://arxiv.org/abs/https://doi.org/10.1080/08927022.2022.2094430}
  {https://doi.org/10.1080/08927022.2022.2094430} \BibitemShut {NoStop}%
\bibitem [{\citenamefont {Niklasson}\ \emph {et~al.}(2007)\citenamefont
  {Niklasson}, \citenamefont {Tymczak},\ and\ \citenamefont
  {Challacombe}}]{ANiklasson07}%
  \BibitemOpen
  \bibfield  {author} {\bibinfo {author} {\bibfnamefont {A.~M.~N.}\
  \bibnamefont {Niklasson}}, \bibinfo {author} {\bibfnamefont {C.~J.}\
  \bibnamefont {Tymczak}}, \ and\ \bibinfo {author} {\bibfnamefont
  {M.}~\bibnamefont {Challacombe}},\ }\href@noop {} {\bibfield  {journal}
  {\bibinfo  {journal} {J. Chem. Phys.}\ }\textbf {\bibinfo {volume} {126}},\
  \bibinfo {pages} {144103} (\bibinfo {year} {2007})}\BibitemShut {NoStop}%
\bibitem [{\citenamefont {Niklasson}(2008)}]{ANiklasson08}%
  \BibitemOpen
  \bibfield  {author} {\bibinfo {author} {\bibfnamefont {A.~M.~N.}\
  \bibnamefont {Niklasson}},\ }\href@noop {} {\bibfield  {journal} {\bibinfo
  {journal} {Phys. Rev. Lett.}\ }\textbf {\bibinfo {volume} {100}},\ \bibinfo
  {pages} {123004} (\bibinfo {year} {2008})}\BibitemShut {NoStop}%
\bibitem [{\citenamefont {Cawkwell}\ and\ \citenamefont
  {Niklasson}(2012)}]{MCawkwell12}%
  \BibitemOpen
  \bibfield  {author} {\bibinfo {author} {\bibfnamefont {M.~J.}\ \bibnamefont
  {Cawkwell}}\ and\ \bibinfo {author} {\bibfnamefont {A.~M.~N.}\ \bibnamefont
  {Niklasson}},\ }\href@noop {} {\bibfield  {journal} {\bibinfo  {journal} {J.
  Chem. Phys.}\ }\textbf {\bibinfo {volume} {137}},\ \bibinfo {pages} {134105}
  (\bibinfo {year} {2012})}\BibitemShut {NoStop}%
\bibitem [{\citenamefont {Hutter}(2012)}]{JHutter12}%
  \BibitemOpen
  \bibfield  {author} {\bibinfo {author} {\bibfnamefont {J.}~\bibnamefont
  {Hutter}},\ }\href@noop {} {\bibfield  {journal} {\bibinfo  {journal} {WIREs
  Comput. Mol. Sci.}\ }\textbf {\bibinfo {volume} {2}},\ \bibinfo {pages} {604}
  (\bibinfo {year} {2012})}\BibitemShut {NoStop}%
\bibitem [{\citenamefont {Lin}\ \emph {et~al.}(2014)\citenamefont {Lin},
  \citenamefont {Lu},\ and\ \citenamefont {Shao}}]{LLin14}%
  \BibitemOpen
  \bibfield  {author} {\bibinfo {author} {\bibfnamefont {L.}~\bibnamefont
  {Lin}}, \bibinfo {author} {\bibfnamefont {J.}~\bibnamefont {Lu}}, \ and\
  \bibinfo {author} {\bibfnamefont {S.}~\bibnamefont {Shao}},\ }\href@noop {}
  {\bibfield  {journal} {\bibinfo  {journal} {Entropy}\ }\textbf {\bibinfo
  {volume} {16}},\ \bibinfo {pages} {110} (\bibinfo {year} {2014})}\BibitemShut
  {NoStop}%
\bibitem [{\citenamefont {Souvatzis}\ and\ \citenamefont
  {Niklasson}(2014)}]{PSouvatzis14}%
  \BibitemOpen
  \bibfield  {author} {\bibinfo {author} {\bibfnamefont {P.}~\bibnamefont
  {Souvatzis}}\ and\ \bibinfo {author} {\bibfnamefont {A.~M.~N.}\ \bibnamefont
  {Niklasson}},\ }\href@noop {} {\bibfield  {journal} {\bibinfo  {journal} {J.
  Chem. Phys.}\ }\textbf {\bibinfo {volume} {140}},\ \bibinfo {pages} {044117}
  (\bibinfo {year} {2014})}\BibitemShut {NoStop}%
\bibitem [{\citenamefont {Niklasson}(2017)}]{ANiklasson17}%
  \BibitemOpen
  \bibfield  {author} {\bibinfo {author} {\bibfnamefont {A.~M.~N.}\
  \bibnamefont {Niklasson}},\ }\href@noop {} {\bibfield  {journal} {\bibinfo
  {journal} {J. Chem. Phys.}\ }\textbf {\bibinfo {volume} {147}},\ \bibinfo
  {pages} {054103} (\bibinfo {year} {2017})}\BibitemShut {NoStop}%
\bibitem [{\citenamefont {Niklasson}(2021{\natexlab{a}})}]{ANiklasson21b}%
  \BibitemOpen
  \bibfield  {author} {\bibinfo {author} {\bibfnamefont {A.~M.~N.}\
  \bibnamefont {Niklasson}},\ }\href@noop {} {\bibfield  {journal} {\bibinfo
  {journal} {Eur. Phys. J. B}\ }\textbf {\bibinfo {volume} {94}},\ \bibinfo
  {pages} {164} (\bibinfo {year} {2021}{\natexlab{a}})}\BibitemShut {NoStop}%
\bibitem [{\citenamefont {Patrick~Henning}(2021)}]{PHenning21}%
  \BibitemOpen
  \bibfield  {author} {\bibinfo {author} {\bibfnamefont {A.~M. N.~N.}\
  \bibnamefont {Patrick~Henning}},\ }\href@noop {} {\bibfield  {journal}
  {\bibinfo  {journal} {Kinetic and Related Models}\ }\textbf {\bibinfo
  {volume} {14}},\ \bibinfo {pages} {303} (\bibinfo {year} {2021})}\BibitemShut
  {NoStop}%
\bibitem [{\citenamefont {Niklasson}(2021{\natexlab{b}})}]{ANiklasson21}%
  \BibitemOpen
  \bibfield  {author} {\bibinfo {author} {\bibfnamefont {A.~M.~N.}\
  \bibnamefont {Niklasson}},\ }\href@noop {} {\bibfield  {journal} {\bibinfo
  {journal} {J. Chem. Phys.}\ }\textbf {\bibinfo {volume} {154}},\ \bibinfo
  {pages} {0000} (\bibinfo {year} {2021}{\natexlab{b}})}\BibitemShut {NoStop}%
\bibitem [{\citenamefont {Marx}\ and\ \citenamefont {Hutter}(2000)}]{DMarx00}%
  \BibitemOpen
  \bibfield  {author} {\bibinfo {author} {\bibfnamefont {D.}~\bibnamefont
  {Marx}}\ and\ \bibinfo {author} {\bibfnamefont {J.}~\bibnamefont {Hutter}},\
  }\enquote {\bibinfo {title} {Modern methods and algorithms of quantum
  chemistry},}\ \ (\bibinfo  {publisher} {ed. J. Grotendorst},\ \bibinfo
  {address} {John von Neumann Institute for Computing, J\"ulich, Germany},\
  \bibinfo {year} {2000})\ \bibinfo {edition} {2nd}\ ed.\BibitemShut {Stop}%
\bibitem [{\citenamefont {Tuckerman}(2010)}]{MTuckerman10}%
  \BibitemOpen
  \bibfield  {author} {\bibinfo {author} {\bibfnamefont {M.~E.}\ \bibnamefont
  {Tuckerman}},\ }\href@noop {} {\emph {\bibinfo {title} {Statistical
  Mechanics: Theory and Molecular Simulation}}}\ (\bibinfo  {publisher} {Oxford
  University Press},\ \bibinfo {address} {New York},\ \bibinfo {year}
  {2010})\BibitemShut {NoStop}%
\bibitem [{\citenamefont {Roothaan}(1951)}]{Roothaan}%
  \BibitemOpen
  \bibfield  {author} {\bibinfo {author} {\bibfnamefont {C.~C.~J.}\
  \bibnamefont {Roothaan}},\ }\href@noop {} {\bibfield  {journal} {\bibinfo
  {journal} {Rev. Mod. Phys.}\ }\textbf {\bibinfo {volume} {23}},\ \bibinfo
  {pages} {69} (\bibinfo {year} {1951})}\BibitemShut {NoStop}%
\bibitem [{\citenamefont {McWeeny}(1959)}]{McWeenyHF}%
  \BibitemOpen
  \bibfield  {author} {\bibinfo {author} {\bibfnamefont {R.}~\bibnamefont
  {McWeeny}},\ }\href {\doibase 10.1103/PhysRev.114.1528} {\bibfield  {journal}
  {\bibinfo  {journal} {Phys. Rev.}\ }\textbf {\bibinfo {volume} {114}},\
  \bibinfo {pages} {1528} (\bibinfo {year} {1959})}\BibitemShut {NoStop}%
\bibitem [{\citenamefont {Mermin}(1963)}]{NMermin63}%
  \BibitemOpen
  \bibfield  {author} {\bibinfo {author} {\bibfnamefont {N.~D.}\ \bibnamefont
  {Mermin}},\ }\href@noop {} {\bibfield  {journal} {\bibinfo  {journal} {Annals
  of Physics}\ }\textbf {\bibinfo {volume} {21}},\ \bibinfo {pages} {99}
  (\bibinfo {year} {1963})}\BibitemShut {NoStop}%
\bibitem [{\citenamefont {Hohenberg}\ and\ \citenamefont {Kohn}(1964)}]{hohen}%
  \BibitemOpen
  \bibfield  {author} {\bibinfo {author} {\bibfnamefont {P.}~\bibnamefont
  {Hohenberg}}\ and\ \bibinfo {author} {\bibfnamefont {W.}~\bibnamefont
  {Kohn}},\ }\href@noop {} {\bibfield  {journal} {\bibinfo  {journal} {Phys.
  Rev.}\ }\textbf {\bibinfo {volume} {136}},\ \bibinfo {pages} {B:864}
  (\bibinfo {year} {1964})}\BibitemShut {NoStop}%
\bibitem [{\citenamefont {Kohn}\ and\ \citenamefont {Sham}(1965)}]{KohnSham65}%
  \BibitemOpen
  \bibfield  {author} {\bibinfo {author} {\bibfnamefont {W.}~\bibnamefont
  {Kohn}}\ and\ \bibinfo {author} {\bibfnamefont {L.~J.}\ \bibnamefont
  {Sham}},\ }\href@noop {} {\bibfield  {journal} {\bibinfo  {journal} {Phys.
  Rev.}\ }\textbf {\bibinfo {volume} {140}},\ \bibinfo {pages} {1133} (\bibinfo
  {year} {1965})}\BibitemShut {NoStop}%
\bibitem [{\citenamefont {Mermin}(1965)}]{NMermin65}%
  \BibitemOpen
  \bibfield  {author} {\bibinfo {author} {\bibfnamefont {N.~D.}\ \bibnamefont
  {Mermin}},\ }\href@noop {} {\bibfield  {journal} {\bibinfo  {journal} {Phys.
  Rev. B}\ }\textbf {\bibinfo {volume} {137}},\ \bibinfo {pages} {A1441}
  (\bibinfo {year} {1965})}\BibitemShut {NoStop}%
\bibitem [{\citenamefont {Parr}\ and\ \citenamefont {Yang}(1989)}]{RParr89}%
  \BibitemOpen
  \bibfield  {author} {\bibinfo {author} {\bibfnamefont {R.~G.}\ \bibnamefont
  {Parr}}\ and\ \bibinfo {author} {\bibfnamefont {W.}~\bibnamefont {Yang}},\
  }\href@noop {} {\emph {\bibinfo {title} {Density-functional theory of atoms
  and molecules}}}\ (\bibinfo  {publisher} {Oxford University Press},\ \bibinfo
  {address} {Oxford},\ \bibinfo {year} {1989})\BibitemShut {NoStop}%
\bibitem [{\citenamefont {Dreizler}\ and\ \citenamefont
  {Gross}(1990)}]{RMDreizler90}%
  \BibitemOpen
  \bibfield  {author} {\bibinfo {author} {\bibfnamefont {R.}~\bibnamefont
  {Dreizler}}\ and\ \bibinfo {author} {\bibfnamefont {K.}~\bibnamefont
  {Gross}},\ }\href@noop {} {\emph {\bibinfo {title} {Density-functional
  theory}}}\ (\bibinfo  {publisher} {Springer Verlag},\ \bibinfo {address}
  {Berlin Heidelberg},\ \bibinfo {year} {1990})\BibitemShut {NoStop}%
\bibitem [{\citenamefont {Engel}\ and\ \citenamefont
  {Dreizler}(2011)}]{EEngel11}%
  \BibitemOpen
  \bibfield  {author} {\bibinfo {author} {\bibfnamefont {E.}~\bibnamefont
  {Engel}}\ and\ \bibinfo {author} {\bibfnamefont {R.}~\bibnamefont
  {Dreizler}},\ }\href@noop {} {\emph {\bibinfo {title} {Density-functional
  theory}}}\ (\bibinfo  {publisher} {Springer Verlag},\ \bibinfo {address}
  {Berlin Heidelberg},\ \bibinfo {year} {2011})\BibitemShut {NoStop}%
\bibitem [{\citenamefont {Remler}\ and\ \citenamefont
  {Madden}(1990)}]{DRemler90}%
  \BibitemOpen
  \bibfield  {author} {\bibinfo {author} {\bibfnamefont {D.~K.}\ \bibnamefont
  {Remler}}\ and\ \bibinfo {author} {\bibfnamefont {P.~A.}\ \bibnamefont
  {Madden}},\ }\href@noop {} {\bibfield  {journal} {\bibinfo  {journal} {Mol.\
  Phys.}\ }\textbf {\bibinfo {volume} {70}},\ \bibinfo {pages} {921} (\bibinfo
  {year} {1990})}\BibitemShut {NoStop}%
\bibitem [{\citenamefont {Pulay}\ and\ \citenamefont
  {Fogarasi}(2004)}]{PPulay04}%
  \BibitemOpen
  \bibfield  {author} {\bibinfo {author} {\bibfnamefont {P.}~\bibnamefont
  {Pulay}}\ and\ \bibinfo {author} {\bibfnamefont {G.}~\bibnamefont
  {Fogarasi}},\ }\href@noop {} {\bibfield  {journal} {\bibinfo  {journal}
  {Chem. Phys. Lett.}\ }\textbf {\bibinfo {volume} {386}},\ \bibinfo {pages}
  {272} (\bibinfo {year} {2004})}\BibitemShut {NoStop}%
\bibitem [{\citenamefont {Herbert}\ and\ \citenamefont
  {Head-Gordon}(2005)}]{JMHerbert05}%
  \BibitemOpen
  \bibfield  {author} {\bibinfo {author} {\bibfnamefont {J.}~\bibnamefont
  {Herbert}}\ and\ \bibinfo {author} {\bibfnamefont {M.}~\bibnamefont
  {Head-Gordon}},\ }\href@noop {} {\bibfield  {journal} {\bibinfo  {journal}
  {Phys. Chem. Chem. Phys.}\ }\textbf {\bibinfo {volume} {7}},\ \bibinfo
  {pages} {3269} (\bibinfo {year} {2005})}\BibitemShut {NoStop}%
\bibitem [{\citenamefont {Niklasson}\ \emph {et~al.}(2006)\citenamefont
  {Niklasson}, \citenamefont {Tymczak},\ and\ \citenamefont
  {Challacombe}}]{ANiklasson06}%
  \BibitemOpen
  \bibfield  {author} {\bibinfo {author} {\bibfnamefont {A.~M.~N.}\
  \bibnamefont {Niklasson}}, \bibinfo {author} {\bibfnamefont {C.~J.}\
  \bibnamefont {Tymczak}}, \ and\ \bibinfo {author} {\bibfnamefont
  {M.}~\bibnamefont {Challacombe}},\ }\href@noop {} {\bibfield  {journal}
  {\bibinfo  {journal} {Phys. Rev. Lett.}\ }\textbf {\bibinfo {volume} {97}},\
  \bibinfo {pages} {123001} (\bibinfo {year} {2006})}\BibitemShut {NoStop}%
\bibitem [{\citenamefont {K\"{u}hne}\ \emph {et~al.}(2007)\citenamefont
  {K\"{u}hne}, \citenamefont {Krack}, \citenamefont {Mohamed},\ and\
  \citenamefont {Parrinello}}]{TDKuhne07}%
  \BibitemOpen
  \bibfield  {author} {\bibinfo {author} {\bibfnamefont {T.~D.}\ \bibnamefont
  {K\"{u}hne}}, \bibinfo {author} {\bibfnamefont {M.}~\bibnamefont {Krack}},
  \bibinfo {author} {\bibfnamefont {F.~R.}\ \bibnamefont {Mohamed}}, \ and\
  \bibinfo {author} {\bibfnamefont {M.}~\bibnamefont {Parrinello}},\
  }\href@noop {} {\bibfield  {journal} {\bibinfo  {journal} {Phys. Rev. Lett.}\
  }\textbf {\bibinfo {volume} {98}},\ \bibinfo {pages} {066401} (\bibinfo
  {year} {2007})}\BibitemShut {NoStop}%
\bibitem [{\citenamefont {Car}\ and\ \citenamefont
  {Parrinello}(1985)}]{RCar85}%
  \BibitemOpen
  \bibfield  {author} {\bibinfo {author} {\bibfnamefont {R.}~\bibnamefont
  {Car}}\ and\ \bibinfo {author} {\bibfnamefont {M.}~\bibnamefont
  {Parrinello}},\ }\href@noop {} {\bibfield  {journal} {\bibinfo  {journal}
  {Phys. Rev. Lett.}\ }\textbf {\bibinfo {volume} {55}},\ \bibinfo {pages}
  {2471} (\bibinfo {year} {1985})}\BibitemShut {NoStop}%
\bibitem [{\citenamefont {Elstner}\ \emph {et~al.}(1998)\citenamefont
  {Elstner}, \citenamefont {Poresag}, \citenamefont {Jungnickel}, \citenamefont
  {Elsner}, \citenamefont {Haugk}, \citenamefont {Frauenheim}, \citenamefont
  {Suhai},\ and\ \citenamefont {Seifert}}]{MElstner98}%
  \BibitemOpen
  \bibfield  {author} {\bibinfo {author} {\bibfnamefont {M.}~\bibnamefont
  {Elstner}}, \bibinfo {author} {\bibfnamefont {D.}~\bibnamefont {Poresag}},
  \bibinfo {author} {\bibfnamefont {G.}~\bibnamefont {Jungnickel}}, \bibinfo
  {author} {\bibfnamefont {J.}~\bibnamefont {Elsner}}, \bibinfo {author}
  {\bibfnamefont {M.}~\bibnamefont {Haugk}}, \bibinfo {author} {\bibfnamefont
  {T.}~\bibnamefont {Frauenheim}}, \bibinfo {author} {\bibfnamefont
  {S.}~\bibnamefont {Suhai}}, \ and\ \bibinfo {author} {\bibfnamefont
  {G.}~\bibnamefont {Seifert}},\ }\href@noop {} {\bibfield  {journal} {\bibinfo
   {journal} {Phys. Rev. B}\ }\textbf {\bibinfo {volume} {58}},\ \bibinfo
  {pages} {7260} (\bibinfo {year} {1998})}\BibitemShut {NoStop}%
\bibitem [{\citenamefont {Finnis}\ \emph {et~al.}(1998)\citenamefont {Finnis},
  \citenamefont {Paxton}, \citenamefont {Methfessel},\ and\ \citenamefont {van
  Schilfgarde}}]{MFinnis98}%
  \BibitemOpen
  \bibfield  {author} {\bibinfo {author} {\bibfnamefont {M.~W.}\ \bibnamefont
  {Finnis}}, \bibinfo {author} {\bibfnamefont {A.~T.}\ \bibnamefont {Paxton}},
  \bibinfo {author} {\bibfnamefont {M.}~\bibnamefont {Methfessel}}, \ and\
  \bibinfo {author} {\bibfnamefont {M.}~\bibnamefont {van Schilfgarde}},\
  }\href@noop {} {\bibfield  {journal} {\bibinfo  {journal} {Phys. Rev. Lett.}\
  }\textbf {\bibinfo {volume} {81}},\ \bibinfo {pages} {5149} (\bibinfo {year}
  {1998})}\BibitemShut {NoStop}%
\bibitem [{\citenamefont {et~al.}(2020)}]{BHourahine20}%
  \BibitemOpen
  \bibfield  {author} {\bibinfo {author} {\bibfnamefont {B.~H.}\ \bibnamefont
  {et~al.}},\ }\href@noop {} {\bibfield  {journal} {\bibinfo  {journal} {J.
  Chem. Phys.}\ }\textbf {\bibinfo {volume} {152}},\ \bibinfo {pages} {124101}
  (\bibinfo {year} {2020})}\BibitemShut {NoStop}%
\bibitem [{\citenamefont {Dewar}\ and\ \citenamefont {Thiel}(1977)}]{MDewar77}%
  \BibitemOpen
  \bibfield  {author} {\bibinfo {author} {\bibfnamefont {M.~J.~S.}\
  \bibnamefont {Dewar}}\ and\ \bibinfo {author} {\bibfnamefont
  {W.}~\bibnamefont {Thiel}},\ }\href@noop {} {\bibfield  {journal} {\bibinfo
  {journal} {Theoret.Chim. Acta}\ }\textbf {\bibinfo {volume} {46}},\ \bibinfo
  {pages} {89} (\bibinfo {year} {1977})}\BibitemShut {NoStop}%
\bibitem [{\citenamefont {Dewar}\ \emph {et~al.}(1985)\citenamefont {Dewar},
  \citenamefont {Zoebisch}, \citenamefont {Healy},\ and\ \citenamefont
  {Stewart}}]{MDewar85}%
  \BibitemOpen
  \bibfield  {author} {\bibinfo {author} {\bibfnamefont {M.~J.~S.}\
  \bibnamefont {Dewar}}, \bibinfo {author} {\bibfnamefont {E.~G.}\ \bibnamefont
  {Zoebisch}}, \bibinfo {author} {\bibfnamefont {E.~F.}\ \bibnamefont {Healy}},
  \ and\ \bibinfo {author} {\bibfnamefont {J.~J.~P.}\ \bibnamefont {Stewart}},\
  }\href@noop {} {\bibfield  {journal} {\bibinfo  {journal} {J. Am. Chem.
  Soc.}\ }\textbf {\bibinfo {volume} {107}},\ \bibinfo {pages} {3902} (\bibinfo
  {year} {1985})}\BibitemShut {NoStop}%
\bibitem [{\citenamefont {Stewart}(2013)}]{JStewart13}%
  \BibitemOpen
  \bibfield  {author} {\bibinfo {author} {\bibfnamefont {J.~J.~P.}\
  \bibnamefont {Stewart}},\ }\href@noop {} {\bibfield  {journal} {\bibinfo
  {journal} {J. Mol. Model.}\ }\textbf {\bibinfo {volume} {19}},\ \bibinfo
  {pages} {1} (\bibinfo {year} {2013})}\BibitemShut {NoStop}%
\bibitem [{\citenamefont {Bannwarth}\ \emph {et~al.}(2018)\citenamefont
  {Bannwarth}, \citenamefont {Ehlert},\ and\ \citenamefont
  {Grimme}}]{CBannwarth18}%
  \BibitemOpen
  \bibfield  {author} {\bibinfo {author} {\bibfnamefont {C.}~\bibnamefont
  {Bannwarth}}, \bibinfo {author} {\bibfnamefont {S.}~\bibnamefont {Ehlert}}, \
  and\ \bibinfo {author} {\bibfnamefont {S.}~\bibnamefont {Grimme}},\
  }\href@noop {} {\bibfield  {journal} {\bibinfo  {journal} {J. Chem. Theory
  Comput.}\ }\textbf {\bibinfo {volume} {15}},\ \bibinfo {pages} {1652}
  (\bibinfo {year} {2018})}\BibitemShut {NoStop}%
\bibitem [{\citenamefont {Dral}\ \emph {et~al.}(2019)\citenamefont {Dral},
  \citenamefont {Wu},\ and\ \citenamefont {Thiel}}]{PDral19}%
  \BibitemOpen
  \bibfield  {author} {\bibinfo {author} {\bibfnamefont {P.~O.}\ \bibnamefont
  {Dral}}, \bibinfo {author} {\bibfnamefont {X.}~\bibnamefont {Wu}}, \ and\
  \bibinfo {author} {\bibfnamefont {W.}~\bibnamefont {Thiel}},\ }\href@noop {}
  {\bibfield  {journal} {\bibinfo  {journal} {J. Chem. Theory Comput.}\
  }\textbf {\bibinfo {volume} {15}},\ \bibinfo {pages} {1743} (\bibinfo {year}
  {2019})}\BibitemShut {NoStop}%
\bibitem [{\citenamefont {Malone}\ \emph {et~al.}(2020)\citenamefont {Malone},
  \citenamefont {Nebgen}, \citenamefont {White}, \citenamefont {Zhang},
  \citenamefont {Song}, \citenamefont {Bjorgaard}, \citenamefont {Sifain},
  \citenamefont {Rodriguez-Hernandez}, \citenamefont {Freixas}, \citenamefont
  {Fernandez-Alberti}, \citenamefont {Roitberg}, \citenamefont {Nelson},\ and\
  \citenamefont {Tretiak}}]{WMalone20}%
  \BibitemOpen
  \bibfield  {author} {\bibinfo {author} {\bibfnamefont {W.}~\bibnamefont
  {Malone}}, \bibinfo {author} {\bibfnamefont {B.}~\bibnamefont {Nebgen}},
  \bibinfo {author} {\bibfnamefont {A.}~\bibnamefont {White}}, \bibinfo
  {author} {\bibfnamefont {Y.}~\bibnamefont {Zhang}}, \bibinfo {author}
  {\bibfnamefont {H.}~\bibnamefont {Song}}, \bibinfo {author} {\bibfnamefont
  {J.~A.}\ \bibnamefont {Bjorgaard}}, \bibinfo {author} {\bibfnamefont {A.~E.}\
  \bibnamefont {Sifain}}, \bibinfo {author} {\bibfnamefont {B.}~\bibnamefont
  {Rodriguez-Hernandez}}, \bibinfo {author} {\bibfnamefont {V.~M.}\
  \bibnamefont {Freixas}}, \bibinfo {author} {\bibfnamefont {S.}~\bibnamefont
  {Fernandez-Alberti}}, \bibinfo {author} {\bibfnamefont {A.~E.}\ \bibnamefont
  {Roitberg}}, \bibinfo {author} {\bibfnamefont {T.~R.}\ \bibnamefont
  {Nelson}}, \ and\ \bibinfo {author} {\bibfnamefont {S.}~\bibnamefont
  {Tretiak}},\ }\href {\doibase 10.1021/acs.jctc.0c00248} {\bibfield  {journal}
  {\bibinfo  {journal} {Journal of Chemical Theory and Computation}\ }\textbf
  {\bibinfo {volume} {16}},\ \bibinfo {pages} {5771} (\bibinfo {year}
  {2020})},\ \bibinfo {note} {pMID: 32635739},\ \Eprint
  {http://arxiv.org/abs/https://doi.org/10.1021/acs.jctc.0c00248}
  {https://doi.org/10.1021/acs.jctc.0c00248} \BibitemShut {NoStop}%
\bibitem [{\citenamefont {Zhou}\ \emph {et~al.}(2020)\citenamefont {Zhou},
  \citenamefont {Nebgen}, \citenamefont {Lubbers}, \citenamefont {Malone},
  \citenamefont {Niklasson},\ and\ \citenamefont {Tretiak}}]{ZGuoqing20}%
  \BibitemOpen
  \bibfield  {author} {\bibinfo {author} {\bibfnamefont {G.}~\bibnamefont
  {Zhou}}, \bibinfo {author} {\bibfnamefont {B.}~\bibnamefont {Nebgen}},
  \bibinfo {author} {\bibfnamefont {N.}~\bibnamefont {Lubbers}}, \bibinfo
  {author} {\bibfnamefont {W.}~\bibnamefont {Malone}}, \bibinfo {author}
  {\bibfnamefont {A.~M.~N.}\ \bibnamefont {Niklasson}}, \ and\ \bibinfo
  {author} {\bibfnamefont {S.}~\bibnamefont {Tretiak}},\ }\href {\doibase
  10.1021/acs.jctc.0c00243} {\bibfield  {journal} {\bibinfo  {journal} {Journal
  of Chemical Theory and Computation}\ }\textbf {\bibinfo {volume} {16}},\
  \bibinfo {pages} {4951} (\bibinfo {year} {2020})},\ \bibinfo {note} {pMID:
  32609513},\ \Eprint
  {http://arxiv.org/abs/https://doi.org/10.1021/acs.jctc.0c00243}
  {https://doi.org/10.1021/acs.jctc.0c00243} \BibitemShut {NoStop}%
\bibitem [{\citenamefont {Bannwarth}\ \emph {et~al.}(2020)\citenamefont
  {Bannwarth}, \citenamefont {Caldeweyher}, \citenamefont {Ehlert},
  \citenamefont {ans P.~Pracht}, \citenamefont {Seibert}, \citenamefont
  {Spicher},\ and\ \citenamefont {Grimme}}]{CBannwarth20}%
  \BibitemOpen
  \bibfield  {author} {\bibinfo {author} {\bibfnamefont {C.}~\bibnamefont
  {Bannwarth}}, \bibinfo {author} {\bibfnamefont {E.}~\bibnamefont
  {Caldeweyher}}, \bibinfo {author} {\bibfnamefont {S.}~\bibnamefont {Ehlert}},
  \bibinfo {author} {\bibfnamefont {A.~H.}\ \bibnamefont {ans P.~Pracht}},
  \bibinfo {author} {\bibfnamefont {J.}~\bibnamefont {Seibert}}, \bibinfo
  {author} {\bibfnamefont {S.}~\bibnamefont {Spicher}}, \ and\ \bibinfo
  {author} {\bibfnamefont {S.}~\bibnamefont {Grimme}},\ }\href@noop {}
  {\bibfield  {journal} {\bibinfo  {journal} {WIREs Comput. Lol. Sci.}\
  }\textbf {\bibinfo {volume} {11}},\ \bibinfo {pages} {1} (\bibinfo {year}
  {2020})}\BibitemShut {NoStop}%
\bibitem [{\citenamefont {Vesely}(1977)}]{FJVesely77}%
  \BibitemOpen
  \bibfield  {author} {\bibinfo {author} {\bibfnamefont {F.~J.}\ \bibnamefont
  {Vesely}},\ }\href@noop {} {\bibfield  {journal} {\bibinfo  {journal} {J.
  Comput. Phys.}\ }\textbf {\bibinfo {volume} {24}},\ \bibinfo {pages} {361}
  (\bibinfo {year} {1977})}\BibitemShut {NoStop}%
\bibitem [{\citenamefont {Sprik}\ and\ \citenamefont {Klein}(1988)}]{MSprik88}%
  \BibitemOpen
  \bibfield  {author} {\bibinfo {author} {\bibfnamefont {M.}~\bibnamefont
  {Sprik}}\ and\ \bibinfo {author} {\bibfnamefont {M.~L.}\ \bibnamefont
  {Klein}},\ }\href {\doibase 10.1063/1.455722} {\bibfield  {journal} {\bibinfo
   {journal} {The Journal of Chemical Physics}\ }\textbf {\bibinfo {volume}
  {89}},\ \bibinfo {pages} {7556} (\bibinfo {year} {1988})},\ \Eprint
  {http://arxiv.org/abs/https://doi.org/10.1063/1.455722}
  {https://doi.org/10.1063/1.455722} \BibitemShut {NoStop}%
\bibitem [{\citenamefont {Mortier}\ \emph {et~al.}(1986)\citenamefont
  {Mortier}, \citenamefont {Ghosh},\ and\ \citenamefont
  {Shankar}}]{WJMortier86}%
  \BibitemOpen
  \bibfield  {author} {\bibinfo {author} {\bibfnamefont {W.~J.}\ \bibnamefont
  {Mortier}}, \bibinfo {author} {\bibfnamefont {S.~K.}\ \bibnamefont {Ghosh}},
  \ and\ \bibinfo {author} {\bibfnamefont {S.}~\bibnamefont {Shankar}},\ }\href
  {\doibase 10.1021/ja00275a013} {\bibfield  {journal} {\bibinfo  {journal}
  {Journal of the American Chemical Society}\ }\textbf {\bibinfo {volume}
  {108}},\ \bibinfo {pages} {4315} (\bibinfo {year} {1986})},\ \Eprint
  {http://arxiv.org/abs/https://doi.org/10.1021/ja00275a013}
  {https://doi.org/10.1021/ja00275a013} \BibitemShut {NoStop}%
\bibitem [{\citenamefont {Rappe}\ and\ \citenamefont {III}(1991)}]{AKRappe91}%
  \BibitemOpen
  \bibfield  {author} {\bibinfo {author} {\bibfnamefont {A.~K.}\ \bibnamefont
  {Rappe}}\ and\ \bibinfo {author} {\bibfnamefont {W.~A.~G.}\ \bibnamefont
  {III}},\ }\href@noop {} {\bibfield  {journal} {\bibinfo  {journal} {J. Phys.
  Chem}\ }\textbf {\bibinfo {volume} {95}},\ \bibinfo {pages} {3358} (\bibinfo
  {year} {1991})}\BibitemShut {NoStop}%
\bibitem [{\citenamefont {Lamoureux}\ and\ \citenamefont
  {Roux}(2003)}]{GLamoureaux03}%
  \BibitemOpen
  \bibfield  {author} {\bibinfo {author} {\bibfnamefont {G.}~\bibnamefont
  {Lamoureux}}\ and\ \bibinfo {author} {\bibfnamefont {B.~T.}\ \bibnamefont
  {Roux}},\ }\href@noop {} {\bibfield  {journal} {\bibinfo  {journal} {J. Chem.
  Phys.}\ }\textbf {\bibinfo {volume} {119}},\ \bibinfo {pages} {3025}
  (\bibinfo {year} {2003})}\BibitemShut {NoStop}%
\bibitem [{\citenamefont {Verstraelen}\ \emph {et~al.}(2013)\citenamefont
  {Verstraelen}, \citenamefont {Ayers}, \citenamefont {Van~Speybroeck},\ and\
  \citenamefont {Waroquier}}]{TVerstraelen13}%
  \BibitemOpen
  \bibfield  {author} {\bibinfo {author} {\bibfnamefont {T.}~\bibnamefont
  {Verstraelen}}, \bibinfo {author} {\bibfnamefont {P.~W.}\ \bibnamefont
  {Ayers}}, \bibinfo {author} {\bibfnamefont {V.}~\bibnamefont
  {Van~Speybroeck}}, \ and\ \bibinfo {author} {\bibfnamefont {M.}~\bibnamefont
  {Waroquier}},\ }\href {\doibase 10.1063/1.4791569} {\bibfield  {journal}
  {\bibinfo  {journal} {The Journal of Chemical Physics}\ }\textbf {\bibinfo
  {volume} {138}},\ \bibinfo {pages} {074108} (\bibinfo {year} {2013})},\
  \Eprint {http://arxiv.org/abs/https://doi.org/10.1063/1.4791569}
  {https://doi.org/10.1063/1.4791569} \BibitemShut {NoStop}%
\bibitem [{\citenamefont {Naserifar}\ \emph {et~al.}(2017)\citenamefont
  {Naserifar}, \citenamefont {Brooks}, \citenamefont {Goddard},\ and\
  \citenamefont {Cvicek}}]{SNaserifar17}%
  \BibitemOpen
  \bibfield  {author} {\bibinfo {author} {\bibfnamefont {S.}~\bibnamefont
  {Naserifar}}, \bibinfo {author} {\bibfnamefont {D.~J.}\ \bibnamefont
  {Brooks}}, \bibinfo {author} {\bibfnamefont {W.~A.}\ \bibnamefont {Goddard}},
  \ and\ \bibinfo {author} {\bibfnamefont {V.}~\bibnamefont {Cvicek}},\ }\href
  {\doibase 10.1063/1.4978891} {\bibfield  {journal} {\bibinfo  {journal} {The
  Journal of Chemical Physics}\ }\textbf {\bibinfo {volume} {146}},\ \bibinfo
  {pages} {124117} (\bibinfo {year} {2017})},\ \Eprint
  {http://arxiv.org/abs/https://doi.org/10.1063/1.4978891}
  {https://doi.org/10.1063/1.4978891} \BibitemShut {NoStop}%
\bibitem [{\citenamefont {York}\ and\ \citenamefont {Yang}(1996)}]{DMYork96}%
  \BibitemOpen
  \bibfield  {author} {\bibinfo {author} {\bibfnamefont {D.~M.}\ \bibnamefont
  {York}}\ and\ \bibinfo {author} {\bibfnamefont {W.}~\bibnamefont {Yang}},\
  }\href {\doibase 10.1063/1.470886} {\bibfield  {journal} {\bibinfo  {journal}
  {The Journal of Chemical Physics}\ }\textbf {\bibinfo {volume} {104}},\
  \bibinfo {pages} {159} (\bibinfo {year} {1996})},\ \Eprint
  {http://arxiv.org/abs/https://doi.org/10.1063/1.470886}
  {https://doi.org/10.1063/1.470886} \BibitemShut {NoStop}%
\bibitem [{\citenamefont {Tabacchi}\ \emph {et~al.}(2002)\citenamefont
  {Tabacchi}, \citenamefont {Mundy}, \citenamefont {Hutter},\ and\
  \citenamefont {Parrinello}}]{GTabacchi02}%
  \BibitemOpen
  \bibfield  {author} {\bibinfo {author} {\bibfnamefont {G.}~\bibnamefont
  {Tabacchi}}, \bibinfo {author} {\bibfnamefont {C.~J.}\ \bibnamefont {Mundy}},
  \bibinfo {author} {\bibfnamefont {J.}~\bibnamefont {Hutter}}, \ and\ \bibinfo
  {author} {\bibfnamefont {M.}~\bibnamefont {Parrinello}},\ }\href@noop {}
  {\bibfield  {journal} {\bibinfo  {journal} {J. Chem. Phys.}\ }\textbf
  {\bibinfo {volume} {117}},\ \bibinfo {pages} {1416} (\bibinfo {year}
  {2002})}\BibitemShut {NoStop}%
\bibitem [{\citenamefont {Harris}(1985)}]{JHarris85}%
  \BibitemOpen
  \bibfield  {author} {\bibinfo {author} {\bibfnamefont {J.}~\bibnamefont
  {Harris}},\ }\href@noop {} {\bibfield  {journal} {\bibinfo  {journal} {Phys.
  Rev. B}\ }\textbf {\bibinfo {volume} {31}},\ \bibinfo {pages} {1770}
  (\bibinfo {year} {1985})}\BibitemShut {NoStop}%
\bibitem [{\citenamefont {Foulkes}\ and\ \citenamefont
  {Haydock}(1989)}]{MFoulkes89}%
  \BibitemOpen
  \bibfield  {author} {\bibinfo {author} {\bibfnamefont {W.~M.~C.}\
  \bibnamefont {Foulkes}}\ and\ \bibinfo {author} {\bibfnamefont
  {R.}~\bibnamefont {Haydock}},\ }\href@noop {} {\bibfield  {journal} {\bibinfo
   {journal} {Phys. Rev. B}\ }\textbf {\bibinfo {volume} {39}},\ \bibinfo
  {pages} {12520} (\bibinfo {year} {1989})}\BibitemShut {NoStop}%
\bibitem [{\citenamefont {Harrison}(1980)}]{WHarrison80}%
  \BibitemOpen
  \bibfield  {author} {\bibinfo {author} {\bibfnamefont {W.~A.}\ \bibnamefont
  {Harrison}},\ }\href@noop {} {\emph {\bibinfo {title} {Electronic structure
  and the properties of solids: the physics of the chemical bond}}}\ (\bibinfo
  {publisher} {Dover},\ \bibinfo {address} {New York},\ \bibinfo {year}
  {1980})\BibitemShut {NoStop}%
\bibitem [{\citenamefont {Porezag}\ \emph {et~al.}(1995)\citenamefont
  {Porezag}, \citenamefont {Frauenheim}, \citenamefont {K\"ohler},
  \citenamefont {Seifert},\ and\ \citenamefont {Kaschner}}]{DPorezag95}%
  \BibitemOpen
  \bibfield  {author} {\bibinfo {author} {\bibfnamefont {D.}~\bibnamefont
  {Porezag}}, \bibinfo {author} {\bibfnamefont {T.}~\bibnamefont {Frauenheim}},
  \bibinfo {author} {\bibfnamefont {T.}~\bibnamefont {K\"ohler}}, \bibinfo
  {author} {\bibfnamefont {G.}~\bibnamefont {Seifert}}, \ and\ \bibinfo
  {author} {\bibfnamefont {R.}~\bibnamefont {Kaschner}},\ }\href {\doibase
  10.1103/PhysRevB.51.12947} {\bibfield  {journal} {\bibinfo  {journal} {Phys.
  Rev. B}\ }\textbf {\bibinfo {volume} {51}},\ \bibinfo {pages} {12947}
  (\bibinfo {year} {1995})}\BibitemShut {NoStop}%
\bibitem [{\citenamefont {Frauenheim}\ \emph {et~al.}(2000)\citenamefont
  {Frauenheim}, \citenamefont {Seifert}, \citenamefont {Elstner}, \citenamefont
  {Hajnal}, \citenamefont {Jungnickel}, \citenamefont {Poresag}, \citenamefont
  {Suhai},\ and\ \citenamefont {Scholz}}]{TFrauenheim00}%
  \BibitemOpen
  \bibfield  {author} {\bibinfo {author} {\bibfnamefont {T.}~\bibnamefont
  {Frauenheim}}, \bibinfo {author} {\bibfnamefont {G.}~\bibnamefont {Seifert}},
  \bibinfo {author} {\bibfnamefont {M.}~\bibnamefont {Elstner}}, \bibinfo
  {author} {\bibfnamefont {Z.}~\bibnamefont {Hajnal}}, \bibinfo {author}
  {\bibfnamefont {G.}~\bibnamefont {Jungnickel}}, \bibinfo {author}
  {\bibfnamefont {D.}~\bibnamefont {Poresag}}, \bibinfo {author} {\bibfnamefont
  {S.}~\bibnamefont {Suhai}}, \ and\ \bibinfo {author} {\bibfnamefont
  {R.}~\bibnamefont {Scholz}},\ }\href@noop {} {\bibfield  {journal} {\bibinfo
  {journal} {Phys. Stat. sol.}\ }\textbf {\bibinfo {volume} {217}},\ \bibinfo
  {pages} {41} (\bibinfo {year} {2000})}\BibitemShut {NoStop}%
\bibitem [{\citenamefont {Koskinen}\ and\ \citenamefont
  {Mäkinen}(2009)}]{PKoskinen09}%
  \BibitemOpen
  \bibfield  {author} {\bibinfo {author} {\bibfnamefont {P.}~\bibnamefont
  {Koskinen}}\ and\ \bibinfo {author} {\bibfnamefont {V.}~\bibnamefont
  {Mäkinen}},\ }\href {\doibase
  https://doi.org/10.1016/j.commatsci.2009.07.013} {\bibfield  {journal}
  {\bibinfo  {journal} {Computational Materials Science}\ }\textbf {\bibinfo
  {volume} {47}},\ \bibinfo {pages} {237 } (\bibinfo {year}
  {2009})}\BibitemShut {NoStop}%
\bibitem [{\citenamefont {Gaus}\ \emph {et~al.}(2011)\citenamefont {Gaus},
  \citenamefont {Cui},\ and\ \citenamefont {Elstner}}]{MGaus11}%
  \BibitemOpen
  \bibfield  {author} {\bibinfo {author} {\bibfnamefont {M.}~\bibnamefont
  {Gaus}}, \bibinfo {author} {\bibfnamefont {Q.}~\bibnamefont {Cui}}, \ and\
  \bibinfo {author} {\bibfnamefont {M.}~\bibnamefont {Elstner}},\ }\href@noop
  {} {\bibfield  {journal} {\bibinfo  {journal} {J, Chem. Theory Comput.}\
  }\textbf {\bibinfo {volume} {7}},\ \bibinfo {pages} {931} (\bibinfo {year}
  {2011})}\BibitemShut {NoStop}%
\bibitem [{\citenamefont {Aradi}\ \emph {et~al.}(2015)\citenamefont {Aradi},
  \citenamefont {Niklasson},\ and\ \citenamefont {Frauenheim}}]{BAradi15}%
  \BibitemOpen
  \bibfield  {author} {\bibinfo {author} {\bibfnamefont {B.}~\bibnamefont
  {Aradi}}, \bibinfo {author} {\bibfnamefont {A.~M.~N.}\ \bibnamefont
  {Niklasson}}, \ and\ \bibinfo {author} {\bibfnamefont {T.}~\bibnamefont
  {Frauenheim}},\ }\href@noop {} {\bibfield  {journal} {\bibinfo  {journal} {J.
  Chem. Theory Comput.}\ }\textbf {\bibinfo {volume} {11}},\ \bibinfo {pages}
  {3357} (\bibinfo {year} {2015})}\BibitemShut {NoStop}%
\bibitem [{Note1()}]{Note1}%
  \BibitemOpen
  \bibinfo {note} {All physically relevant electron densities determined by
  anti-symmetric electron wavefunctions \cite
  {RMDreizler90,EEngel11}}\BibitemShut {NoStop}%
\bibitem [{\citenamefont {Pittalis}\ \emph {et~al.}(2011)\citenamefont
  {Pittalis}, \citenamefont {Proetto}, \citenamefont {Floris}, \citenamefont
  {Sanna}, \citenamefont {Bersier}, \citenamefont {Burke},\ and\ \citenamefont
  {Gross}}]{SPittalis11}%
  \BibitemOpen
  \bibfield  {author} {\bibinfo {author} {\bibfnamefont {S.}~\bibnamefont
  {Pittalis}}, \bibinfo {author} {\bibfnamefont {C.~R.}\ \bibnamefont
  {Proetto}}, \bibinfo {author} {\bibfnamefont {A.}~\bibnamefont {Floris}},
  \bibinfo {author} {\bibfnamefont {A.}~\bibnamefont {Sanna}}, \bibinfo
  {author} {\bibfnamefont {C.}~\bibnamefont {Bersier}}, \bibinfo {author}
  {\bibfnamefont {K.}~\bibnamefont {Burke}}, \ and\ \bibinfo {author}
  {\bibfnamefont {E.~K.~U.}\ \bibnamefont {Gross}},\ }\href {\doibase
  10.1103/PhysRevLett.107.163001} {\bibfield  {journal} {\bibinfo  {journal}
  {Phys. Rev. Lett.}\ }\textbf {\bibinfo {volume} {107}},\ \bibinfo {pages}
  {163001} (\bibinfo {year} {2011})}\BibitemShut {NoStop}%
\bibitem [{\citenamefont {Pribram-Jones}\ \emph {et~al.}(2014)\citenamefont
  {Pribram-Jones}, \citenamefont {Pittalis}, \citenamefont {Gross},\ and\
  \citenamefont {Burke}}]{APJones14}%
  \BibitemOpen
  \bibfield  {author} {\bibinfo {author} {\bibfnamefont {A.}~\bibnamefont
  {Pribram-Jones}}, \bibinfo {author} {\bibfnamefont {S.}~\bibnamefont
  {Pittalis}}, \bibinfo {author} {\bibfnamefont {E.~K.~U.}\ \bibnamefont
  {Gross}}, \ and\ \bibinfo {author} {\bibfnamefont {K.}~\bibnamefont
  {Burke}},\ }in\ \href@noop {} {\emph {\bibinfo {booktitle} {Frontiers and
  Challenges in Warm Dense Matter}}},\ \bibinfo {editor} {edited by\ \bibinfo
  {editor} {\bibfnamefont {F.}~\bibnamefont {Graziani}}, \bibinfo {editor}
  {\bibfnamefont {M.~P.}\ \bibnamefont {Desjarlais}}, \bibinfo {editor}
  {\bibfnamefont {R.}~\bibnamefont {Redmer}}, \ and\ \bibinfo {editor}
  {\bibfnamefont {S.~B.}\ \bibnamefont {Trickey}}}\ (\bibinfo  {publisher}
  {Springer International Publishing},\ \bibinfo {address} {Cham},\ \bibinfo
  {year} {2014})\ pp.\ \bibinfo {pages} {25--60}\BibitemShut {NoStop}%
\bibitem [{\citenamefont {Heitler}\ and\ \citenamefont
  {London}(1927)}]{WHeitler27}%
  \BibitemOpen
  \bibfield  {author} {\bibinfo {author} {\bibfnamefont {W.}~\bibnamefont
  {Heitler}}\ and\ \bibinfo {author} {\bibfnamefont {F.}~\bibnamefont
  {London}},\ }\href@noop {} {\bibfield  {journal} {\bibinfo  {journal} {Z.\
  Phys.}\ }\textbf {\bibinfo {volume} {44}},\ \bibinfo {pages} {455} (\bibinfo
  {year} {1927})}\BibitemShut {NoStop}%
\bibitem [{\citenamefont {Born}\ and\ \citenamefont
  {Oppenheimer}(1927)}]{MBorn27}%
  \BibitemOpen
  \bibfield  {author} {\bibinfo {author} {\bibfnamefont {M.}~\bibnamefont
  {Born}}\ and\ \bibinfo {author} {\bibfnamefont {R.}~\bibnamefont
  {Oppenheimer}},\ }\href@noop {} {\bibfield  {journal} {\bibinfo  {journal}
  {Ann.\ Phys.}\ }\textbf {\bibinfo {volume} {389}},\ \bibinfo {pages} {475}
  (\bibinfo {year} {1927})}\BibitemShut {NoStop}%
\bibitem [{\citenamefont {Tuckerman}(2002)}]{MTuckerman02}%
  \BibitemOpen
  \bibfield  {author} {\bibinfo {author} {\bibfnamefont {M.~E.}\ \bibnamefont
  {Tuckerman}},\ }\href@noop {} {\bibfield  {journal} {\bibinfo  {journal} {J.
  Phys.: Conden. Matter}\ }\textbf {\bibinfo {volume} {14}},\ \bibinfo {pages}
  {1297} (\bibinfo {year} {2002})}\BibitemShut {NoStop}%
\bibitem [{Note2()}]{Note2}%
  \BibitemOpen
  \bibinfo {note} {Notice that exact ground state, $\rho _{\protect \rm
  min}({\protect \bf r})$, and its approximate ground state solution, $\rho
  \approx \rho _{\protect \rm min}$, depend on ${\protect \bf R}$, i.e.\ $\rho
  _{\protect \rm min}({\protect \bf r}) \equiv \rho _{\protect \rm
  min}({\protect \bf R},{\protect \bf r})$, but we have dropped the explicit
  ${\protect \bf R}$-dependencies in our simplified notation}\BibitemShut
  {NoStop}%
\bibitem [{Note3()}]{Note3}%
  \BibitemOpen
  \bibinfo {note} {If $n^{(0)} = \rho _{\protect \rm min}$ then $\rho
  _{\protect \rm min}[n^{(0)}] = n^{(0)}$ and ${\protect \cal U}^{(0)} =
  U$.}\BibitemShut {Stop}%
\bibitem [{\citenamefont {Niklasson}(2020{\natexlab{a}})}]{ANiklasson20}%
  \BibitemOpen
  \bibfield  {author} {\bibinfo {author} {\bibfnamefont {A.~M.~N.}\
  \bibnamefont {Niklasson}},\ }\href@noop {} {\bibfield  {journal} {\bibinfo
  {journal} {J. Chem. Phys.}\ }\textbf {\bibinfo {volume} {152}},\ \bibinfo
  {pages} {104103} (\bibinfo {year} {2020}{\natexlab{a}})}\BibitemShut
  {NoStop}%
\bibitem [{\citenamefont {Niklasson}(2020{\natexlab{b}})}]{ANiklasson20b}%
  \BibitemOpen
  \bibfield  {author} {\bibinfo {author} {\bibfnamefont {A.~M.~N.}\
  \bibnamefont {Niklasson}},\ }\href@noop {} {\bibfield  {journal} {\bibinfo
  {journal} {J. Chem. Theory Comput.}\ }\textbf {\bibinfo {volume} {16}},\
  \bibinfo {pages} {3628} (\bibinfo {year} {2020}{\natexlab{b}})}\BibitemShut
  {NoStop}%
\bibitem [{\citenamefont {Das}\ and\ \citenamefont {Gavini}(2022)}]{VGavini22}%
  \BibitemOpen
  \bibfield  {author} {\bibinfo {author} {\bibfnamefont {S.}~\bibnamefont
  {Das}}\ and\ \bibinfo {author} {\bibfnamefont {V.}~\bibnamefont {Gavini}},\
  }\href {\doibase 10.48550/ARXIV.2211.07894} {\enquote {\bibinfo {title}
  {Accelerating self-consistent field iterations in kohn-sham density
  functional theory using a low rank approximation of the dielectric matrix},}\
  } (\bibinfo {year} {2022})\BibitemShut {NoStop}%
\bibitem [{\citenamefont {Negre}\ \emph {et~al.}(2022)\citenamefont {Negre},
  \citenamefont {Wall},\ and\ \citenamefont {Niklasson}}]{CNegre22}%
  \BibitemOpen
  \bibfield  {author} {\bibinfo {author} {\bibfnamefont {C.~F.~A.}\
  \bibnamefont {Negre}}, \bibinfo {author} {\bibfnamefont {M.~E.}\ \bibnamefont
  {Wall}}, \ and\ \bibinfo {author} {\bibfnamefont {A.~M.~N.}\ \bibnamefont
  {Niklasson}},\ }\href {\doibase 10.48550/ARXIV.2212.01997} {\enquote
  {\bibinfo {title} {Graph-based quantum response theory and shadow
  born-oppenheimer molecular dynamics},}\ } (\bibinfo {year}
  {2022})\BibitemShut {NoStop}%
\bibitem [{\citenamefont {Steneteg}\ \emph {et~al.}(2010)\citenamefont
  {Steneteg}, \citenamefont {Abrikosov}, \citenamefont {Weber},\ and\
  \citenamefont {Niklasson}}]{PSteneteg10}%
  \BibitemOpen
  \bibfield  {author} {\bibinfo {author} {\bibfnamefont {P.}~\bibnamefont
  {Steneteg}}, \bibinfo {author} {\bibfnamefont {I.~A.}\ \bibnamefont
  {Abrikosov}}, \bibinfo {author} {\bibfnamefont {V.}~\bibnamefont {Weber}}, \
  and\ \bibinfo {author} {\bibfnamefont {A.~M.~N.}\ \bibnamefont {Niklasson}},\
  }\href@noop {} {\bibfield  {journal} {\bibinfo  {journal} {Phys. Rev. B}\
  }\textbf {\bibinfo {volume} {82}},\ \bibinfo {pages} {075110} (\bibinfo
  {year} {2010})}\BibitemShut {NoStop}%
\bibitem [{\citenamefont {Arita}\ \emph {et~al.}(2014)\citenamefont {Arita},
  \citenamefont {Bowler},\ and\ \citenamefont {Miyazaki}}]{MArita14}%
  \BibitemOpen
  \bibfield  {author} {\bibinfo {author} {\bibfnamefont {M.}~\bibnamefont
  {Arita}}, \bibinfo {author} {\bibfnamefont {D.~R.}\ \bibnamefont {Bowler}}, \
  and\ \bibinfo {author} {\bibfnamefont {T.}~\bibnamefont {Miyazaki}},\
  }\href@noop {} {\bibfield  {journal} {\bibinfo  {journal} {J. Chem. Theory
  Comput.}\ }\textbf {\bibinfo {volume} {10}},\ \bibinfo {pages} {5419}
  (\bibinfo {year} {2014})}\BibitemShut {NoStop}%
\bibitem [{\citenamefont {Peters}\ \emph {et~al.}(2017)\citenamefont {Peters},
  \citenamefont {Kussmann},\ and\ \citenamefont {Ochsenfeld}}]{LDMPeters17}%
  \BibitemOpen
  \bibfield  {author} {\bibinfo {author} {\bibfnamefont {L.~D.~M.}\
  \bibnamefont {Peters}}, \bibinfo {author} {\bibfnamefont {J.}~\bibnamefont
  {Kussmann}}, \ and\ \bibinfo {author} {\bibfnamefont {C.}~\bibnamefont
  {Ochsenfeld}},\ }\href@noop {} {\bibfield  {journal} {\bibinfo  {journal} {J.
  Chem. Theory Comput.}\ }\textbf {\bibinfo {volume} {13}},\ \bibinfo {pages}
  {5479} (\bibinfo {year} {2017})}\BibitemShut {NoStop}%
\bibitem [{\citenamefont {Niklasson}\ \emph {et~al.}(2009)\citenamefont
  {Niklasson}, \citenamefont {Steneteg}, \citenamefont {Odell}, \citenamefont
  {Bock}, \citenamefont {Challacombe}, \citenamefont {Tymczak}, \citenamefont
  {Holmstrom}, \citenamefont {Zheng},\ and\ \citenamefont
  {Weber}}]{ANiklasson09}%
  \BibitemOpen
  \bibfield  {author} {\bibinfo {author} {\bibfnamefont {A.~M.~N.}\
  \bibnamefont {Niklasson}}, \bibinfo {author} {\bibfnamefont {P.}~\bibnamefont
  {Steneteg}}, \bibinfo {author} {\bibfnamefont {A.}~\bibnamefont {Odell}},
  \bibinfo {author} {\bibfnamefont {N.}~\bibnamefont {Bock}}, \bibinfo {author}
  {\bibfnamefont {M.}~\bibnamefont {Challacombe}}, \bibinfo {author}
  {\bibfnamefont {C.~J.}\ \bibnamefont {Tymczak}}, \bibinfo {author}
  {\bibfnamefont {E.}~\bibnamefont {Holmstrom}}, \bibinfo {author}
  {\bibfnamefont {G.}~\bibnamefont {Zheng}}, \ and\ \bibinfo {author}
  {\bibfnamefont {V.}~\bibnamefont {Weber}},\ }\href@noop {} {\bibfield
  {journal} {\bibinfo  {journal} {J. Chem. Phys.}\ }\textbf {\bibinfo {volume}
  {130}},\ \bibinfo {pages} {214109} (\bibinfo {year} {2009})}\BibitemShut
  {NoStop}%
\bibitem [{\citenamefont {Zheng}\ \emph {et~al.}(2011)\citenamefont {Zheng},
  \citenamefont {Niklasson},\ and\ \citenamefont {Karplus}}]{GZheng11}%
  \BibitemOpen
  \bibfield  {author} {\bibinfo {author} {\bibfnamefont {G.}~\bibnamefont
  {Zheng}}, \bibinfo {author} {\bibfnamefont {A.~M.~N.}\ \bibnamefont
  {Niklasson}}, \ and\ \bibinfo {author} {\bibfnamefont {M.}~\bibnamefont
  {Karplus}},\ }\href@noop {} {\bibfield  {journal} {\bibinfo  {journal} {J.
  Chem. Phys.}\ }\textbf {\bibinfo {volume} {135}},\ \bibinfo {pages} {044122}
  (\bibinfo {year} {2011})}\BibitemShut {NoStop}%
\bibitem [{\citenamefont {Odell}\ \emph {et~al.}(2009)\citenamefont {Odell},
  \citenamefont {Delin}, \citenamefont {Johansson}, \citenamefont {Bock},
  \citenamefont {Challacombe},\ and\ \citenamefont {Niklasson}}]{AOdell09}%
  \BibitemOpen
  \bibfield  {author} {\bibinfo {author} {\bibfnamefont {A.}~\bibnamefont
  {Odell}}, \bibinfo {author} {\bibfnamefont {A.}~\bibnamefont {Delin}},
  \bibinfo {author} {\bibfnamefont {B.}~\bibnamefont {Johansson}}, \bibinfo
  {author} {\bibfnamefont {N.}~\bibnamefont {Bock}}, \bibinfo {author}
  {\bibfnamefont {M.}~\bibnamefont {Challacombe}}, \ and\ \bibinfo {author}
  {\bibfnamefont {A.~M.~N.}\ \bibnamefont {Niklasson}},\ }\href@noop {}
  {\bibfield  {journal} {\bibinfo  {journal} {J. Chem. Phys.}\ }\textbf
  {\bibinfo {volume} {131}},\ \bibinfo {pages} {244106} (\bibinfo {year}
  {2009})}\BibitemShut {NoStop}%
\bibitem [{\citenamefont {Odell}\ \emph {et~al.}(2011)\citenamefont {Odell},
  \citenamefont {Delin}, \citenamefont {Johansson}, \citenamefont {Cawkwell},\
  and\ \citenamefont {Niklasson}}]{AOdell11}%
  \BibitemOpen
  \bibfield  {author} {\bibinfo {author} {\bibfnamefont {A.}~\bibnamefont
  {Odell}}, \bibinfo {author} {\bibfnamefont {A.}~\bibnamefont {Delin}},
  \bibinfo {author} {\bibfnamefont {B.}~\bibnamefont {Johansson}}, \bibinfo
  {author} {\bibfnamefont {M.~J.}\ \bibnamefont {Cawkwell}}, \ and\ \bibinfo
  {author} {\bibfnamefont {A.~M.~N.}\ \bibnamefont {Niklasson}},\ }\href@noop
  {} {\bibfield  {journal} {\bibinfo  {journal} {J. Chem. Phys.}\ }\textbf
  {\bibinfo {volume} {135}},\ \bibinfo {pages} {224105} (\bibinfo {year}
  {2011})}\BibitemShut {NoStop}%
\bibitem [{\citenamefont {Leven}\ and\ \citenamefont
  {Head-Gordon}(2019)}]{ILeven19}%
  \BibitemOpen
  \bibfield  {author} {\bibinfo {author} {\bibfnamefont {I.}~\bibnamefont
  {Leven}}\ and\ \bibinfo {author} {\bibfnamefont {T.}~\bibnamefont
  {Head-Gordon}},\ }\href@noop {} {\bibfield  {journal} {\bibinfo  {journal}
  {Phys. Chem. Chem. Phys.}\ }\textbf {\bibinfo {volume} {21}},\ \bibinfo
  {pages} {18652} (\bibinfo {year} {2019})}\BibitemShut {NoStop}%
\bibitem [{\citenamefont {An}\ \emph {et~al.}(2020)\citenamefont {An},
  \citenamefont {Cheng}, \citenamefont {Head-Gordon}, \citenamefont {Lin},\
  and\ \citenamefont {Lu}}]{DAn20}%
  \BibitemOpen
  \bibfield  {author} {\bibinfo {author} {\bibfnamefont {D.}~\bibnamefont
  {An}}, \bibinfo {author} {\bibfnamefont {S.~Y.}\ \bibnamefont {Cheng}},
  \bibinfo {author} {\bibfnamefont {T.}~\bibnamefont {Head-Gordon}}, \bibinfo
  {author} {\bibfnamefont {L.}~\bibnamefont {Lin}}, \ and\ \bibinfo {author}
  {\bibfnamefont {J.}~\bibnamefont {Lu}},\ }\href@noop {} {\enquote {\bibinfo
  {title} {Convergence of stochastic-extended lagrangian molecular dynamics
  method for polarizable force field simulation},}\ } (\bibinfo {year}
  {2020}),\ \Eprint {http://arxiv.org/abs/1904.12082} {arXiv:1904.12082
  [math.NA]} \BibitemShut {NoStop}%
\bibitem [{\citenamefont {Niklasson}\ \emph {et~al.}(2015)\citenamefont
  {Niklasson}, \citenamefont {Cawkwell}, \citenamefont {Rubensson},\ and\
  \citenamefont {Rudberg}}]{ANiklasson15}%
  \BibitemOpen
  \bibfield  {author} {\bibinfo {author} {\bibfnamefont {A.~M.~N.}\
  \bibnamefont {Niklasson}}, \bibinfo {author} {\bibfnamefont {M.~J.}\
  \bibnamefont {Cawkwell}}, \bibinfo {author} {\bibfnamefont {E.~H.}\
  \bibnamefont {Rubensson}}, \ and\ \bibinfo {author} {\bibfnamefont
  {E.}~\bibnamefont {Rudberg}},\ }\href@noop {} {\bibfield  {journal} {\bibinfo
   {journal} {Phys. Rev. E}\ }\textbf {\bibinfo {volume} {92}},\ \bibinfo
  {pages} {063301} (\bibinfo {year} {2015})}\BibitemShut {NoStop}%
\bibitem [{\citenamefont {Nishimoto}(2017)}]{YNishimoto17}%
  \BibitemOpen
  \bibfield  {author} {\bibinfo {author} {\bibfnamefont {Y.}~\bibnamefont
  {Nishimoto}},\ }\href@noop {} {\bibfield  {journal} {\bibinfo  {journal} {J.
  Chem. Phys.}\ }\textbf {\bibinfo {volume} {146}},\ \bibinfo {pages} {084101}
  (\bibinfo {year} {2017})}\BibitemShut {NoStop}%
\bibitem [{\citenamefont {Niklasson}\ and\ \citenamefont
  {Cawkwell}(2014)}]{ANiklasson14}%
  \BibitemOpen
  \bibfield  {author} {\bibinfo {author} {\bibfnamefont {A.~M.~N.}\
  \bibnamefont {Niklasson}}\ and\ \bibinfo {author} {\bibfnamefont
  {M.}~\bibnamefont {Cawkwell}},\ }\href@noop {} {\bibfield  {journal}
  {\bibinfo  {journal} {J. Chem. Phys.}\ }\textbf {\bibinfo {volume} {141}},\
  \bibinfo {pages} {164123} (\bibinfo {year} {2014})}\BibitemShut {NoStop}%
\bibitem [{\citenamefont {Feynman}(1939)}]{RPFeynman39}%
  \BibitemOpen
  \bibfield  {author} {\bibinfo {author} {\bibfnamefont {R.~P.}\ \bibnamefont
  {Feynman}},\ }\href {\doibase 10.1103/PhysRev.56.340} {\bibfield  {journal}
  {\bibinfo  {journal} {Phys. Rev.}\ }\textbf {\bibinfo {volume} {56}},\
  \bibinfo {pages} {340} (\bibinfo {year} {1939})}\BibitemShut {NoStop}%
\bibitem [{\citenamefont {Cawkwell}\ and\ \citenamefont
  {et~al.}(2010)}]{LATTE}%
  \BibitemOpen
  \bibfield  {author} {\bibinfo {author} {\bibfnamefont {M.~J.}\ \bibnamefont
  {Cawkwell}}\ and\ \bibinfo {author} {\bibnamefont {et~al.}},\ }\href@noop {}
  {\enquote {\bibinfo {title} {{\sc LATTE}},}\ } (\bibinfo {year} {2010}),\
  \bibinfo {note} {\mbox{L}os Alamos National Laboratory (LA- CC-10004),
  http://www.github.com/lanl/latte}\BibitemShut {NoStop}%
\bibitem [{\citenamefont {Krishnapryian}\ \emph {et~al.}(2017)\citenamefont
  {Krishnapryian}, \citenamefont {Yang}, \citenamefont {Niklasson},\ and\
  \citenamefont {Cawkwell}}]{AKrishnapriyan17}%
  \BibitemOpen
  \bibfield  {author} {\bibinfo {author} {\bibfnamefont {A.}~\bibnamefont
  {Krishnapryian}}, \bibinfo {author} {\bibfnamefont {P.}~\bibnamefont {Yang}},
  \bibinfo {author} {\bibfnamefont {A.~M.~N.}\ \bibnamefont {Niklasson}}, \
  and\ \bibinfo {author} {\bibfnamefont {M.~J.}\ \bibnamefont {Cawkwell}},\
  }\href@noop {} {\bibfield  {journal} {\bibinfo  {journal} {J. Chem. Theory
  Comput.}\ }\textbf {\bibinfo {volume} {13}},\ \bibinfo {pages} {6191}
  (\bibinfo {year} {2017})}\BibitemShut {NoStop}%
\end{thebibliography}%

\end{document}